\documentclass[journal]{IEEEtran}
\ifCLASSINFOpdf
\else
\fi
\usepackage{cite}
\usepackage{amsmath,amssymb,amsfonts}
\usepackage{algorithmic}
\usepackage{graphicx}
\usepackage{textcomp}
\usepackage{xcolor}
\usepackage{multirow}
\usepackage{booktabs}
\usepackage{threeparttable}
\usepackage{color}
\usepackage{tabularx}
\usepackage{amsmath}
\usepackage{amssymb}
\usepackage{graphicx}
\usepackage{pifont}
\usepackage{diagbox}

\usepackage[ruled]{algorithm2e}
\usepackage{amssymb,amsmath,caption}
\usepackage[hang,flushmargin]{footmisc}

\makeatletter
\newcommand{\algorithmfootnote}[2][\footnotesize]{%
  \let\old@algocf@finish\@algocf@finish
  \def\@algocf@finish{\old@algocf@finish
    \leavevmode\rlap{\begin{minipage}{\linewidth}
    #1#2
    \end{minipage}}%
  }%
}
\makeatother

\begin{document}
%
\title{Tree-Structured Data Clustering-Driven Neural Network for Intra Prediction in Video Coding}
%
%
%

\author{Hengyu~Man,
        Xiaopeng~Fan,~\IEEEmembership{Senior~Member,~IEEE,}
        Ruiqin~Xiong,~\IEEEmembership{Senior~Member,~IEEE,}
        and~Debin~Zhao,~\IEEEmembership{Member,~IEEE}
\thanks{H. Man, X. Fan, and D. Zhao are with the Department of Computer Science and Technology, Harbin Institute of Technology, Harbin 150001, China, and also with Peng Cheng Laboratory,
Shenzhen 518055, China (e-mail: manhengyu@hotmail.com, fxp@hit.edu.cn, dbzhao@hit.edu.cn).}%
\thanks{R. Xiong is with the Institute of Digital Media,
School of Electronic Engineering and Computer Science, Peking University, Beijing 100871, China (e-mail: rqxiong@pku.edu.cn).}%
}

%
%

\markboth{SUBMITTED TO IEEE TRANSACTION ON IMAGE PROCESSING}%
{Shell \MakeLowercase{\textit{et al.}}: Bare Demo of IEEEtran.cls for IEEE Journals}
%



\maketitle

\begin{abstract}
As a crucial part of video compression, intra prediction utilizes local information of images to eliminate the redundancy in spatial domain. In both the High Efficiency Video Coding (H.265/HEVC) and Versatile Video Coding (H.266/VVC), multiple directional prediction modes are employed to find the texture trend of each block and then the prediction is made based on reference samples in the selected direction. Recently, the intra prediction schemes based on neural networks have achieved great success. In these methods, the networks are trained and applied to intra prediction to assist the directional prediction modes. In this paper, we propose a novel tree-structured data clustering-driven neural network (dubbed TreeNet) for intra prediction, which builds the networks and clusters the training data in a tree-structured manner. Specifically, in each network split and training process of TreeNet, every parent network on a leaf node is split into two child networks by adding or subtracting Gaussian random noise. Then a data clustering-driven training is applied to train the two derived child networks using the clustered training data of their parent. To test the performance, TreeNet is integrated into VVC and HEVC to combine with or replace the directional prediction modes. In addition, a fast termination strategy is proposed to accelerate the search of TreeNet. The experimental results demonstrate that TreeNet with the fast termination can reach an average of 2.8\% Bj$\phi$ntegaard distortion rate (BD-rate) improvement (up to 8.1\%) and 4.9\% BD-rate improvement (up to 8.2\%) over VVC (VTM-4.0) and HEVC (HM-16.9) with all intra configuration, respectively.
\end{abstract}

\begin{IEEEkeywords}
Intra prediction, neural networks, network split, data clustering, tree structure.
\end{IEEEkeywords}

%
\IEEEpeerreviewmaketitle

\section{Introduction}
%
%
%
%
\IEEEPARstart{A}{s} the state of the art video coding standard, H.266/VVC \cite{b1} developed by the Joint Video Expert team (JVET) could achieve approximately 50\% bitrate reduction for equivalent perceptual quality compared with its predecessor H.265/HEVC \cite{b2} \cite{b3} with computational complexity increase as a trade-off \cite{b4}. Especially, the bitrate saving provided by intra coding of VVC is 25.1\% on average, and up to 29.3\% over HEVC \cite{VVCintra}. The improvement is mainly achieved by the more flexible block partition, the increased number of angular prediction modes, and additional advanced prediction techniques.

As a crucial part of intra prediction, the granularity of angular prediction modes can greatly influence the intra prediction performance. In HEVC, the number of angular prediction modes is 33. By adding a new mode between every two neighboring angular prediction modes of HEVC, the number of the angular prediction modes in VVC is increased to 65 for square blocks \cite{VVCintra}. The increased angular prediction modes enable VVC to describe the directional patterns in local areas more precisely and thus give more reasonable prediction results. 

Although more angular prediction modes can make intra prediction more accurate, it is still challenging to predict blocks with complex patterns, especially for blocks without obvious directional features. This is because the angular prediction modes in VVC and HEVC are all designed manually and almost equidistantly distributed with a certain direction. What's more, for each mode, only a few reference pixels in a specific direction can be accessed and only simple linear interpolation is performed for prediction.

Recently, neural network-based video coding has been developed rapidly. For the hybrid video coding framework, deep tools are integrated into every single component \cite{b6}, for example, intra prediction \cite{b7} -\cite{PSRNN}, inter prediction \cite{b14}, \cite{b15}, in-loop filtering \cite{b16} -\cite{b18}, entropy coding \cite{b19}, \cite{b20}, and post-processing \cite{b21}, \cite{b22}. For inter prediction, Yan \textit{et al.} \cite{b14} proposed a fractional-pixel reference generation convolutional neural network (CNN) for both uni-directional and bi-directional motion compensation. To improve the bi-prediction performance, Zhao \textit{et al.} \cite{b15} designed an enhanced bi-prediction scheme based on CNN to generate predicted blocks in a non-linear fashion to improve the coding performance. For in-loop filtering, neural networks either act as additional tools to improve the performance of filtering \cite{b16}, \cite{b17}, or replace the deblocking filter and the sample adaptive offset \cite{b18}. Neural networks could also be adopted as entropy coding tools to estimate the probabilities of predefined syntax elements, e.g., the quantized coefficient \cite{b19} and the transform index\cite{b20}. On the decoder side, neural network-based post-processing tools are applied to improve reconstruction quality \cite{b21}, \cite{b22}.

In the existing neural network-based intra predictions, the networks either directly generate the prediction pixels \cite{b7}, \cite{b11} - \cite{PSRNN} or enhance the prediction quality \cite{b9}, \cite{b10} using available reference samples. In \cite{b7}, block-context pairs are separated into two clusters according to a directional context-block relationship to train two neural networks individually. In \cite{TMM}, multiple network modes are trained and each corresponds to a certain range of direction prediction modes. In \cite{b11} - \cite{b13}, more network modes (up to 35) are designed for blocks of different sizes, and the network architecture is further simplified. In \cite{b9} and \cite{b10}, the well-trained CNNs are added after the intra prediction module to refine the prediction.

Motivated by the successful practice of progressively refined angular prediction modes in intra prediction, in this paper, a tree-structured data clustering-driven neural network (TreeNet) is proposed for intra prediction, which builds the networks and clusters the training data in a tree-structured manner rather than the parallel networks as in \cite{b7} and \cite{b11} - \cite{TMM}. Specifically, in each network split and training process of TreeNet, every parent network on a leaf node is split into two child networks by adding or subtracting Gaussian random noise. Then a K-means-like \cite{b23} data clustering-driven training \cite{b16} is applied to train the two derived child networks using the clustered training data of their parent. In addition, a fast termination strategy is proposed to accelerate the search of TreeNet.

The main contributions of this work are summarized as follows:

1) A tree-structured data clustering-driven neural network (dubbed TreeNet) is proposed for intra prediction, which builds the networks and clusters the training data in a tree-structured manner. TreeNet is integrated into VVC and HEVC to combine with or replace the directional prediction modes.

2) A network split strategy is designed to split each parent network on a leaf node into two child networks by adding or subtracting Gaussian random noise. A data clustering-driven training is applied to train the two derived child networks using the clustered training data of their parent.

3) A fast termination strategy is proposed to accelerate the search of TreeNet.

The rest of the paper is organized as follows. Section II provides a brief overview of related works, including intra prediction in HEVC and VVC, and neural network-based intra predictions. The proposed TreeNet is detailed in Section III. In Section IV, the experimental results and more analyses are presented. Finally, the conclusion and future works are provided in Section V. 
\section{Related Works}
In this section, the intra predictions in HEVC and VVC are firstly presented. Then the related neural network-based intra predictions are reviewed.
\subsection{Intra Prediction in HEVC and VVC}
In H.265/HEVC, the quadtree-based block partition is adopted for intra prediction and transformation \cite{b3}. Each picture is divided into disjunct square coding tree units (CTUs) of the same size, each of which serves as the root of a block partition quadtree structure. Along with the coding tree structure, the CTUs can be subdivided into coding units (CUs). A CU can be further split along the coding tree structure into smaller prediction units (PUs) and transform units (TUs). In VVC, the concepts of CU, PU, and TU are replaced by CU. A more advanced block partition mechanism with embedded multi-type trees (MTT), including quadtree (QT), binary tree (BT), and ternary tree (TT), is adopted \cite{b26}.

\begin{figure}[t!]
\centerline{\includegraphics[width=9cm, height=7.8cm]{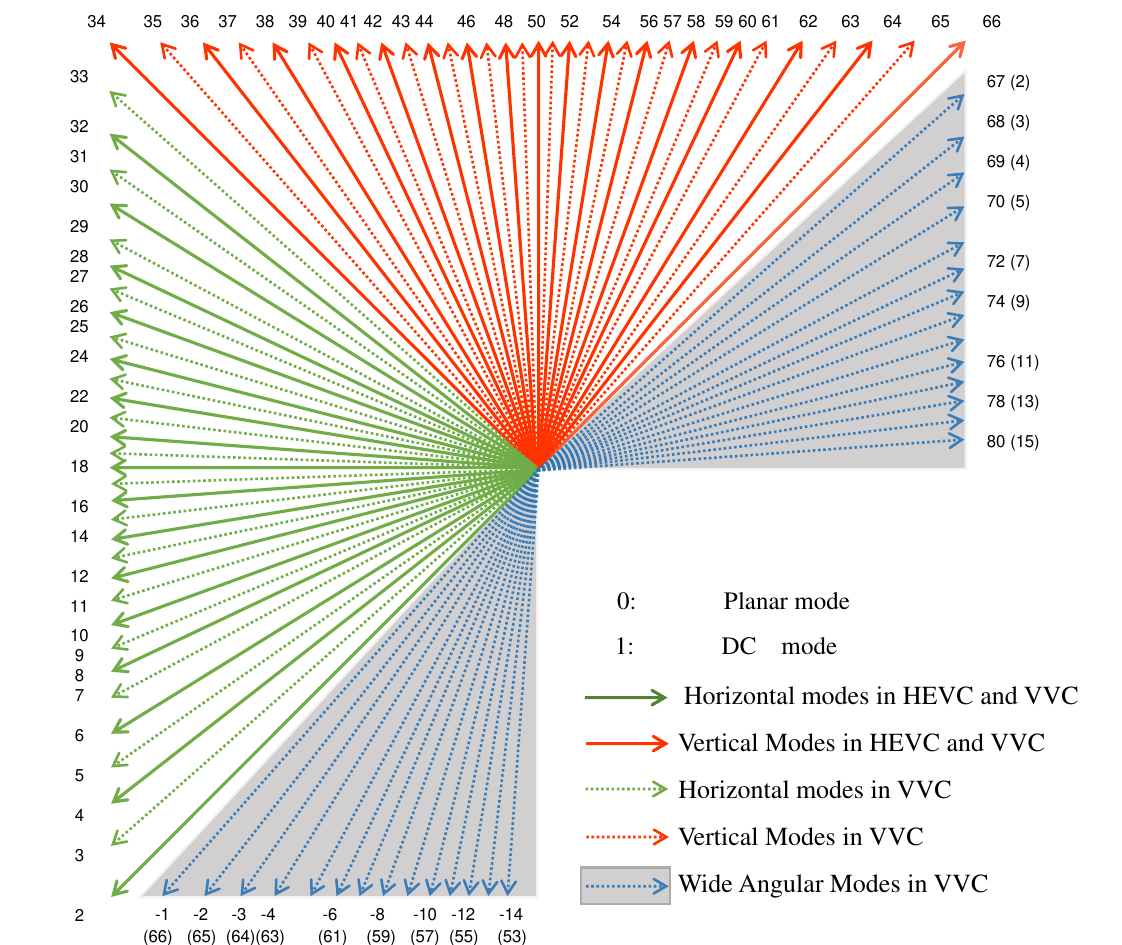}}
\caption{The directional prediction modes in HEVC and VVC. The red and green solid lines represent the vertical modes and the horizontal modes that exist in both HEVC and VVC. Modes represented by dotted lines only exists in VVC. The blue dotted lines in the shadow represent the wide angular modes in VVC. Mode 0 is planar mode and mode 1 is DC mode.}
\label{fig1}
\end{figure}

In HEVC, there are 35 directional prediction modes as shown in Fig. \ref{fig1}, including planar mode, DC mode, and 33 angular prediction modes. Planar mode is designed for areas with gradually changing contents. DC mode aims to predict smooth textures, and the predicted sample values are populated with a constant value representing the average of all surrounding reference samples. Angular prediction modes are designed to model structures with directional edges. To better adapt to the diverse contents and different block sizes, the number of angular prediction modes is increased to 65 in VVC by adding a new mode between every two neighboring angular prediction modes of HEVC as shown in Fig. \ref{fig1}. What's more, aiming at the problem of asymmetric distribution of direction prediction in the non-square block, VVC adopts wide angular intra prediction (WAIP) \cite{b27} in which 14 angles using prediction from the shorter side are replaced by more extreme angles using prediction from the longer side, bringing the total number of intra directions supported in VVC to 93.

In addition to the increased directional prediction modes, multiple additional advanced prediction techniques are designed in VVC to make better use of surrounding reference samples, such as matrix-based intra prediction (MIP) \cite{b12}, position-dependent prediction combination (PDPC) \cite{b28}, multiple reference lines (MRL) \cite{b29}, and cross-component linear model (CCLM) \cite{b30}. In PDPC, filtering with spatially varying weights is applied to blocks that use Planar mode, DC mode, and certain angular prediction modes. To benefit the prediction for sharp content, MRL is designed to use the non-adjacent reference line which is one, two or four lines away from the current block for prediction rather than only using the nearest line as in traditional directional prediction. CCLM is designed for the prediction of chroma components. The chroma components can be predicted from the collocated reconstructed luma samples by linear models whose parameters are derived from adjacent reconstructed samples.

\subsection{Neural Network-Based Intra Prediction}
Recently, neural network-based intra predictions in video coding have been developed rapidly and achieved impressive performance. In \cite{b7}, a fully connected network-based intra prediction (IPFCN) is proposed to learn an end-to-end mapping from neighboring reconstructed pixels to the target block. The neural networks are fed by 8 reference lines above and on the left side of the target block. Two variants of IPFCN: IPFCN-S and IPFCN-D are designed. IPFCN-S has a single network that is trained utilizing an unclassified training data, while IPFCN-D has two networks that are trained by the data from the non-angular prediction modes (DC and Planar) and the other angular prediction modes separately. IPFCN-S and IPFCN-D could achieve 2.9\% and 3.4\% BD-rate improvement over HEVC, respectively.

In \cite{b11}, different numbers of networks (up to 35) are designed for blocks of different sizes. All neural networks for a given block size share all layers but the last one. The network mode signaling is based on a sorted mode list, which is produced by another fully connected network. The bitrate saving in \cite{b11} is around 3.0\% over HEVC. Compared with IPFCN, the network architecture is simplified. In a later version \cite{b12}, each network has only one layer without non-linear activation, i.e. each network is reduced to a matrix plus a bias vector. Only one row and one column from the reconstructed picture are used as reference pixels, and only half of the pixels in the target block are generated by the selected network. The remaining half is generated by linear interpolation. This matrix-based intra prediction (MIP) has been adopted by VVC with a bitrate saving of 0.79\% under all-intra configuration. In \cite{TMM}, multiple network modes (NM) are trained and each corresponds to a certain range of the directional prediction modes. A maximum 6.3\% bitrate reduction is achieved by NM compared with HM-16.9 with 8$\times$8 constraint. In \cite{b13}, a novel training loss function that reflects properties of the residual quantization and coding stages is designed by applying the $L1$-norm and a sigmoid-function to the prediction residual in DCT domain. It could achieve a bitrate saving of 3.79\% compared with VTM 1.0. Different from the fully connected structure applied in above-mentioned works, a progressive spatial recurrent neural network (PS-RNN) which consists of three spatial recurrent units is designed in \cite{PSRNN}, and the prediction is progressively generated by passing information from preceding contents to blocks to be encoded. PS-RNN achieves on average 2.7\% bitrate reduction over HEVC.

Neural network could also be utilized for prediction result refinement. In \cite{b9}, a convolutional neural network-based intra prediction is proposed. The block size is constrained as 8$\times$8. After HEVC intra prediction, the target block, along with its three nearest reconstruction blocks, forms a 16$\times$16 block as the input of CNN. The output of the neural network is a residual block of the same size as the input. By subtracting the generated residual block from the original HEVC prediction result, the prediction quality can be enhanced. MSCNN proposed in \cite{b10} further improves this method by increasing the number of supported block sizes. In addition, the multi-scale feature extraction is proposed to take advantage of feature maps in different scales to improve the performance. MSCNN could achieve an average of 3.4\% bitrate saving with all intra configuration over HEVC.

In this paper, we propose a tree-structured data clustering-driven neural
network (TreeNet) for intra prediction. TreeNet builds the networks and clusters the training data in a tree-structured manner, rather than the parallel networks as in \cite{b7} and \cite{b11} - \cite{TMM}.
\section{The Framework of TreeNet}
In this section, the network structure in TreeNet is first introduced. Then the construction of TreeNet and the data clustering-driven training are presented. After that, the fast termination strategy for the search of TreeNet is described. Finally, the details of how to integrate TreeNet into the VVC and HEVC are provided.

\subsection{Network Structure in TreeNet}
\begin{figure}[t!]
\centerline{\includegraphics[width=9cm, height=8.3cm]{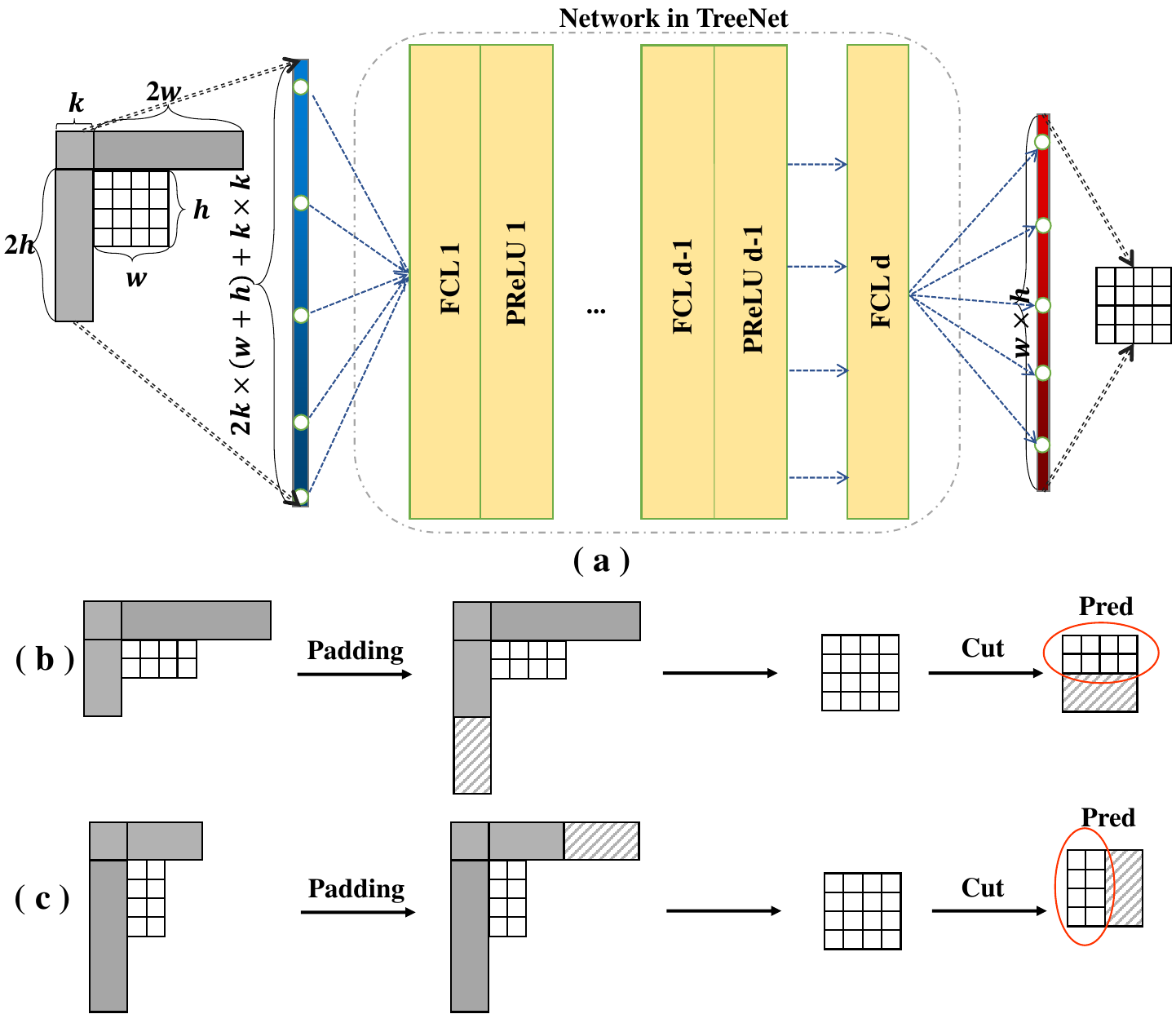}}
\caption{ (a) The structure of a network in TreeNet. (b,c) The prediction of non-square blocks. }
\label{fig2}
\end{figure}

The network structure in TreeNet is shown in Fig. \ref{fig2} (a). The network is composed of four parts: the input layer, the output layer, the fully connected layers, and the non-linear activation layers. The network input is the reconstructed reference samples surrounding the target block. To utilize more spatial contextual information for more accurate prediction, the multi-reference-line scheme \cite{b31} is employed, thus the network input is a $2k\times\left ( w+h \right )+k^{2}$ vector, where $w$ and $h$ are the width and height of the target block and $k$ is the number of reference lines. In TreeNet, $w = h$. If some required reference pixels are not available, the vacancies are filled in by a padding mechanism which is similar to the method used in HEVC \cite{b32}. The network output is an $w\times h$ vector, corresponding to the predicted pixels of the target block in raster order. Since overmuch non-square shapes are supported in VVC, we transform non-square shapes to square shapes as follows. Firstly, the shorter reference line is padded to the same length as the longer one. Then the prediction is obtained after cutting off the redundant part of the network output. The prediction of non-square blocks is shown in Figs. \ref{fig2} (b) and (c). 

All linear layers of the network in TreeNet are fully connected, which is convenient for the network split. Except for the last fully connected layer, all fully connected layers are with the same output dimension. In addition, except for the last fully connected layer, each fully connected layer is followed by a non-linear activation layer. If the depth of the neural network is represented by $d$, the output of the $i_{th}$ layer can be calculated as follows:
\begin{equation}
Y_{i}=
\left\{
\begin{array}{lr}
\sigma (W_{i}\times Y_{i-1}+ B_{i}),\qquad 1\leq i < d\\
W_{i}\times Y_{i-1}+ B_{i}, \qquad \qquad i = d,
\end{array}
\right.
\end{equation}
where $W_{i}$ and $B_{i}$ represent the weights and bias, respectively. $Y_{0}$ and $Y_{d}$ are the input and output of the network. $\sigma$ represents the non-linear activation function. In this paper, Parametric Rectified Linear Unit (PReLU) \cite{b33} is selected as the activation function.

In TreeNet, the number of reference lines $k$ is set as 8, and the network depth $d$ is set as 4, as recommended in \cite{b7}. Different networks are designed for square blocks of different sizes. The layer dimensions of networks are set to be 512, 1024, 1024, 2048, 2048 for 4$\times$4, 8$\times$8, 16$\times$16, 32$\times$32, and 64$\times$64 blocks,respectively.
\subsection{TreeNet Construction}
\begin{figure}[t!]
\centerline{\includegraphics[width=9.4cm, height=8.7cm]{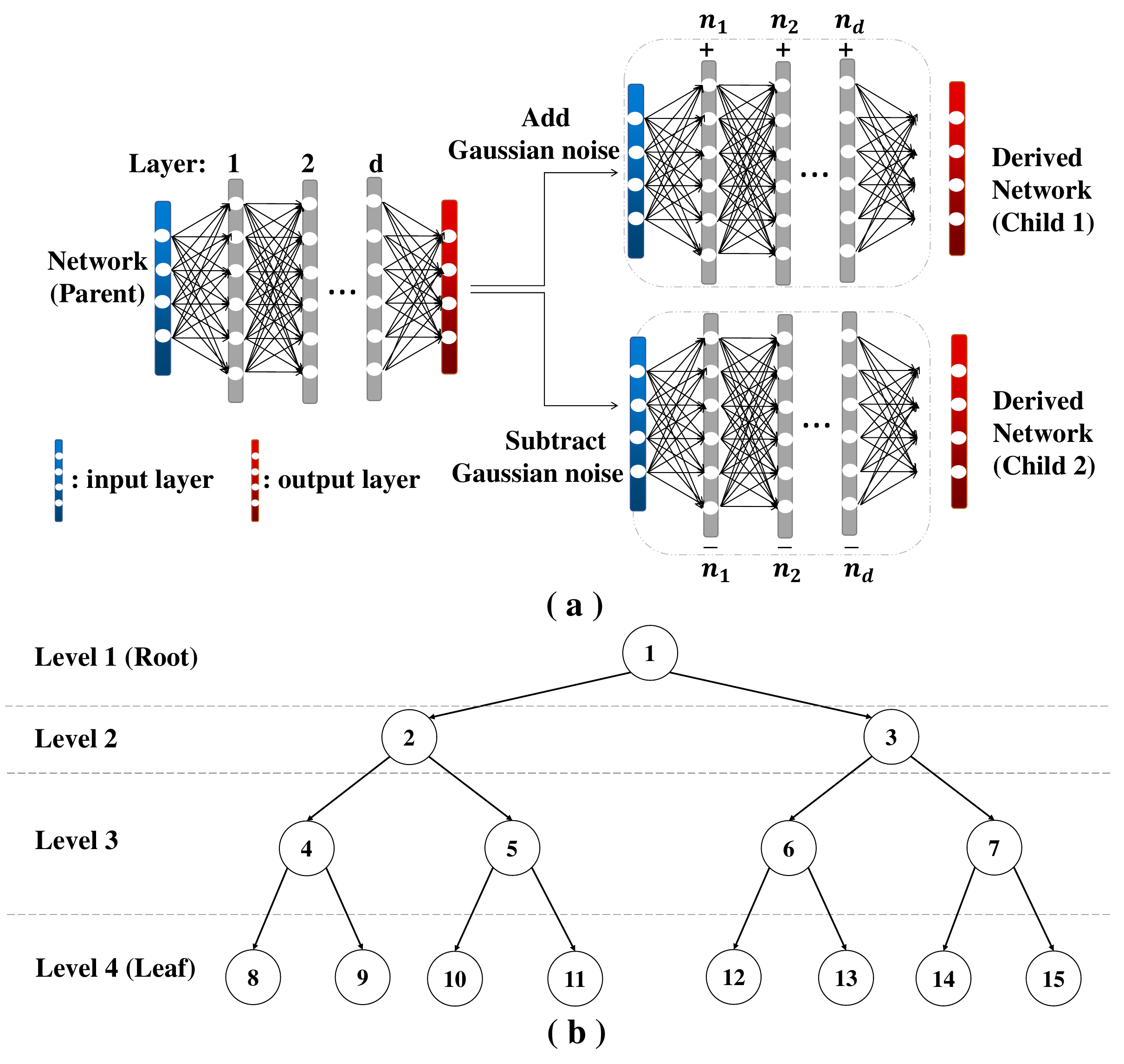}}
\caption{(a) Splitting one parent network into two child networks. (b) The construction of TreeNet with $Depth$=4. \textcircled{i} represent the $i_{th}$ network in TreeNet.}
\label{fig4}
\end{figure}

In traditional Linde-Buzo-Gray (LBG) data clustering algorithm \cite{b34}, a perturbation is produced and applied as an addition or subtraction term for splitting one clustering centroid into two. Inspired by \cite{b34}, we design a network split method for TreeNet. In each network split, every parent network on a leaf node is split into two networks by adding or subtracting Gaussian random noise to its each layer. The variance of the introduced Gaussian random noise is decided by the weights of each layer so that the mapping from the input to the output in the subsequent activation function won't be changed. The two derived child networks have the same structure as their parent network. Fig. \ref{fig4} (a) is an example of splitting one parent network into two child networks, where $n_{i}$ represents the Gaussian random noise introduced to the $i_{th}$ layer.

The networks in TreeNet are built in a tree-structured manner. An example of constructing TreeNet with $Depth$=4 is shown in Fig. \ref{fig4} (b). TreeNet is a perfect binary tree network and the number of networks at level $l$ is $2^{l-1}$. The total number of networks in TreeNet with $Depth$=$L$ is $2^{L}-1$. After each network split, a data clustering-driven training is applied to train the every two derived child networks using the clustered training data of their parent. 

As TreeNet goes deeper, the encoder complexity and memory burden will increase exponentially, and the overhead to indicate the networks will also increase.  If the network split can't bring enough prediction improvement, the network split will be terminated.

\subsection{Data Clustering-Driven Training of TreeNet}
\begin{figure}[t!]
\centerline{\includegraphics[width=9.6cm, height=7.8cm]{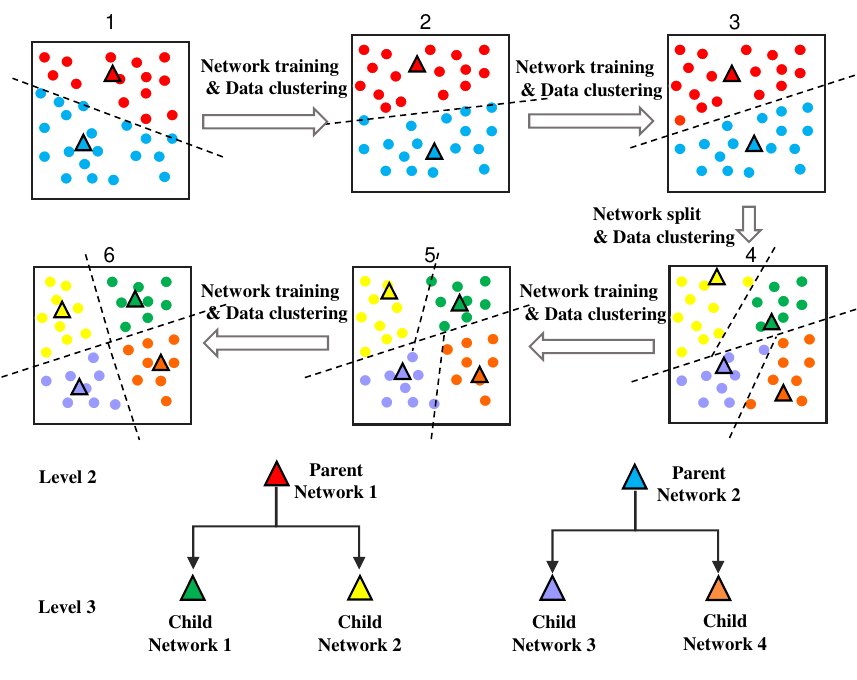}}
\caption{An example of the data clustering and the network training. The triangle filled with a specific color represents the networks, and the circles filled with the same color represent the corresponding training data.}
\label{xx}
\end{figure}

Inspired by the K-means-like training strategy in \cite{b16}, a data clustering-driven training is applied to train every two derived child networks using the clustered training data of their parent. Fig. \ref{xx} is an example of the data clustering and the network training. 

\textit{1) Clustering and Training}: In each iteration of clustering and training, the training data of a parent network is first clustered according to the recovery qualities of the two derived child networks. In this paper, the squared error between original pixels and the prediction is used to measure the recovery quality. Then the clustered training data are fed into the corresponding derived child networks for training. Given a collection of $M$ training sample pairs, the loss function is formulated as:
\begin{equation}
L\left ( \Theta  \right )=\frac{1}{M}\sum_{m=1}^{M}\left \| \mathcal{F}\left ( x_{m}|\Theta  \right ) - y_{m} \right \|_{2}^{2}+\gamma \left \|W   \right \|_{2}^{2},
\end{equation}
where  $\mathcal{F}\left ( x_{m}|\Theta  \right )$ represents the network output and $y_{m}$ represents the ground truth. $\Theta$ is the parameter set. $\gamma$ represents the weight of the $L2$ regularization term, and it is set to be $10^{-4}$ in the experiments. $W$ is the weight of fully connected layers. The optimization algorithm to update all parameters is the stochastic gradient descent ($SGD$) with a momentum term \cite{b35}. The momentum of $SGD$ is set as 0.9.

After several iterations of clustering and training, the training loss gradually converges, which indicates that the clustering centers no longer move. Then the data clustering-driven training will stop.

\textit{2) Training Details}: Firstly, the parameter set $\Theta=\left\{ W,B,A \right\}$ is initialized. The weight $W$ is randomly generated from a Gaussian distribution with a mean of 0 and a standard deviation of 1. The bias $B$ is initialized as 0. The scale factor $A$ in PReLU layers is initialized as 0.25. Then a root network of TreeNet is pre-trained for 80 epochs. The base learning rate is set to 0.1 as recommended in \cite{b7} and \cite{b10}. The network is trained with a fixed learning rate for 20 epochs and then uses the exponentially decaying learning for another 20 epochs until its convergence. The child networks are trained with exponentially decaying learning rates from 0.01 to 0.00001 for 40 epochs. The learning rate settings are listed in Table \ref{table1}.

\begin{table}[t!]
\center
\caption{Setting of Learning Rate}
\renewcommand\arraystretch{1.2}
\setlength{\tabcolsep}{3.5mm}
\scalebox{1}{
\begin{tabular}{c|c|c|c}
\hline
\hline
              & Num of Epochs &  Learning  Rate            & Step \\ \hline
Pre-train      & 80                                                       &  0.1-0.0001     & 20    \\ \hline
Recursive train     & 40                                                       &  0.01-0.00001   & 10    \\ \hline \hline
\end{tabular}
\label{table1}}
\end{table}

\begin{algorithm}[t]
	\renewcommand{\algorithmicrequire}{\textbf{Input:}}
	\renewcommand{\algorithmicensure}{\textbf{Output:}}
	\algorithmfootnote{* $N_{l}[i]$ is the $i_{th}$ network at level $l$.}
	\caption{The fast termination strategy}
	\label{alg:1}
	\begin{algorithmic}[0]
		\REQUIRE TreeNet with $Depth=L$\\
		\qquad\ Lagrange multiplier: $\lambda_{QP}$\\
		\qquad\ $L$-shape reference: $r$, Original pixels: $t$
		\ENSURE Optimal Level: $L_{opt}$, \\
		        \qquad\ \ \ Optimal Network Index: $I_{opt}$\\
		\STATE $I_{opt}$ = 1
	    \FOR{$l = 1$ to $L-1$}
	        \STATE $P \ \ \,= N_{l}[I_{opt}]$ 
	        \STATE $ C_{1} \ \,= N_{l+1}[2 \times I_{opt} - 1] $ , $ C_{2} \ = N_{l+1}[2 \times I_{opt}] $
	        \STATE $J_{ P }\ \, = D(P(r)\ ,t) + \lambda_{QP} \times R_{P}$
	        \STATE $J_{ C_{1} } \, = D(C_{1}(r),t) + \lambda_{QP} \times R_{C_{1}}$
	        \STATE $J_{ C_{2} } \, = D(C_{2}(r),t) + \lambda_{QP} \times R_{C_{2}}$
	        \STATE $\Delta J_{ 1 } = J_{ C_{1} } - J_{ P }$
	        \STATE $\Delta J_{ 2 } = J_{ C_{2} } - J_{ P }$
	        \IF{$\Delta J_{ 1 } \geqslant 0$ and $\Delta J_{ 2 } \geqslant 0 $}
	        \STATE $L_{opt} = l$
		    \STATE \textbf{return} $L_{opt},I_{opt}$
		    \ELSE
		    \IF{$J_{ C_{1} } < J_{ C_{2} }$}
		    \STATE $I_{opt}= I_{opt} \times 2 - 1$
		    \ELSE
		    \STATE $I_{opt}= I_{opt} \times 2$
		    \ENDIF
		    \ENDIF
		\ENDFOR
		\STATE $L_{opt} = L$
		\STATE \textbf{return} $L_{opt},I_{opt}$ 
		\end{algorithmic} 
\end{algorithm}

\subsection{Fast Termination Strategy}
After TreeNet is well-trained, it can be used for intra prediction. However, the full search of TreeNet is very time consuming. Since the training of derived child networks only use the clustered training data from their parent network, the predictions between a parent network and its two child networks are correlated. Therefore, a fast termination strategy for TreeNet is proposed by exploring the correlation, which is shown in Algorithm 1. The rate-distortion (R-D) cost of a network can be expressed as: 
\begin{gather}
J = D + \lambda_{QP}\times R
\end{gather}
where $J$, $D$, and $R$ represent the R-D cost, the sum of squared error (SSE), and costed bits, respectively. $\lambda_{QP}$ is the Lagrange multiplier decided by a quantization parameter (QP). The R-D cost difference $\Delta J$ between a child network and its parent network can be calculated as:
\begin{gather}
\begin{split}
\Delta J &=\left (  D_{C} + \lambda_{QP}\times R_{C}\right ) - \left (D_{P} + \lambda_{QP}\times R_{P}\right )\\
&= \left (D_{C} - D_{P}\right )  +  \lambda_{QP}\times\Delta R\\
&= \Delta D + \lambda_{QP}\times\Delta R
\end{split}
\end{gather}
where the subscripts: $C$ and $P$ represent the child network and the parent network. If $\Delta J_{ 1 }$ and $\Delta J_{ 2 }$ are not smaller than 0, indicating that the parent network is better than both the two derived child networks, the search of TreeNet will be early terminated. Otherwise, the derived child network with less R-D cost will be searched, while the other child network is ignored. The fast termination strategy could help save the encoding running time greatly because at most $2\times L - 1$ networks are searched in TreeNet with $Depth$=$L$ compared with $2^{L}-1$ networks for the full search.

\subsection{Integrating TreeNet in VVC and HEVC}
To test the performance of TreeNet, we integrate TreeNet into VVC and HEVC reference softwares. In VVC, the best prediction mode is decided at the CU level. Fig. 5 shows an example of integrating TreeNet into the VVC framework. RDO means the rate-distortion optimization \cite{b37}. Similar to MRL, TreeNet doesn’t work with Intra Sub-Partitions (ISP) coding mode \cite{ISP}. A new flag $S$ has to be coded and transmitted in the bitstream to indicate whether TreeNet is used or not. In HEVC, the best prediction mode is decided at the PU level. The integration of TreeNet in HEVC is exactly the same as in VVC.

\begin{figure}[t!]
\centerline{\includegraphics[width=9.5cm, height=4.5cm]{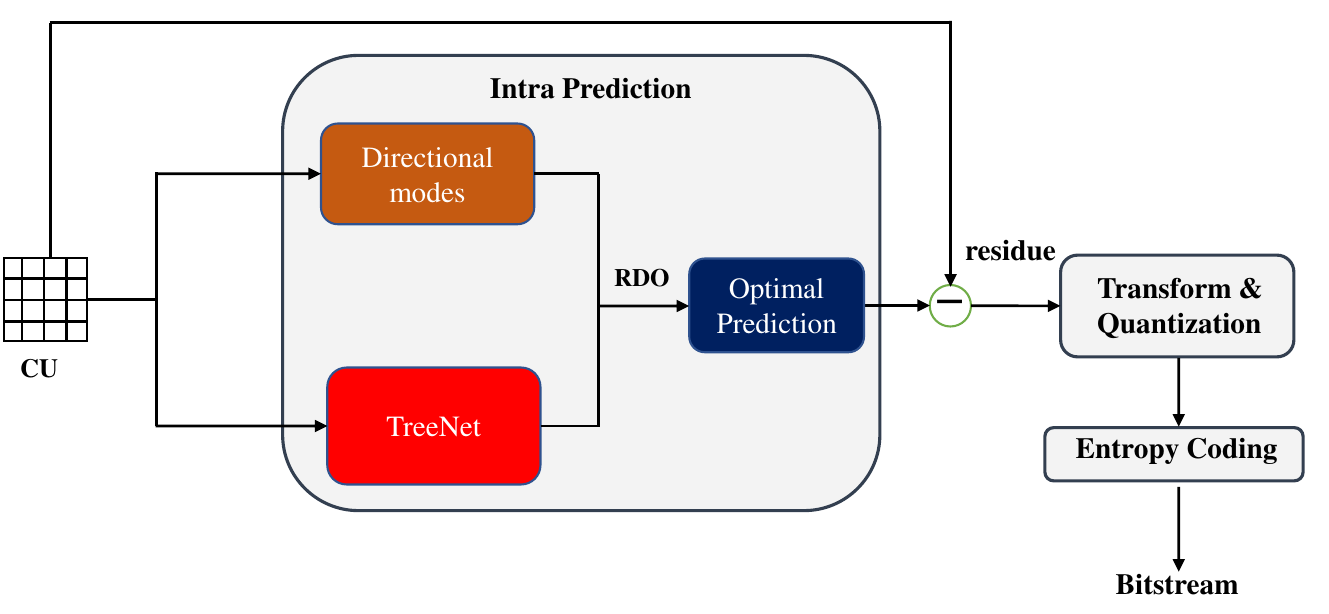}}
\caption{An example of integrating TreeNet into the VVC framework.}
\label{fig8}
\end{figure}

\begin{table}[t!]
\center
\caption{The Codeword of Each TreeNet Mode at Each Level}
\renewcommand\arraystretch{1.15}
\setlength{\tabcolsep}{3.6mm}
\scalebox{1}{           
\begin{tabular}{c|c|cccccc}
\hline \hline
\multirow{2}{*}{\textbf{Level}} & \multirow{2}{*}{\textbf{Mode}} & \multicolumn{6}{c}{\textbf{Binarization}}                                                                                                                                            \\ \cline{3-8} 
                                &                                & \multicolumn{1}{c|}{\textbf{$p$}} & \multicolumn{1}{c|}{\textbf{$c$}} & \multicolumn{1}{c|}{\textbf{$p$}} & \multicolumn{1}{c|}{\textbf{$c$}} & \multicolumn{1}{c|}{\textbf{$p$}} & \textbf{$c$} \\ \hline
\textbf{1}                      & \textbf{1}                     & \multicolumn{1}{c|}{0}          & \multicolumn{1}{c|}{\diagbox[width=3.3em,height=1.1em,dir=SW]{}{}}           & \multicolumn{1}{c|}{\diagbox[width=3.3em,height=1.1em,dir=SW]{}{}}           & \multicolumn{1}{c|}{\diagbox[width=3.3em,height=1.1em,dir=SW]{}{}}           & \multicolumn{1}{c|}{\diagbox[width=3.3em,height=1.1em,dir=SW]{}{}}           &    \diagbox[width=3.3em,height=1.1em,dir=SW]{}{}        \\ \hline
\multirow{2}{*}{\textbf{2}}     & \textbf{2}                     & \multicolumn{1}{c|}{1}          & \multicolumn{1}{c|}{0}          & \multicolumn{1}{c|}{0}          & \multicolumn{1}{c|}{\diagbox[width=3.3em,height=1.1em,dir=SW]{}{}}           & \multicolumn{1}{c|}{\diagbox[width=3.3em,height=1.1em,dir=SW]{}{}}           &  \diagbox[width=3.3em,height=1.1em,dir=SW]{}{}          \\ \cline{2-8} 
                                & \textbf{3}                     & \multicolumn{1}{c|}{1}          & \multicolumn{1}{c|}{1}          & \multicolumn{1}{c|}{0}          & \multicolumn{1}{c|}{\diagbox[width=3.3em,height=1.1em,dir=SW]{}{}}           & \multicolumn{1}{c|}{\diagbox[width=3.3em,height=1.1em,dir=SW]{}{}}           &       \diagbox[width=3.3em,height=1.1em,dir=SW]{}{}     \\ \hline
\multirow{4}{*}{\textbf{3}}     & \textbf{4}                     & \multicolumn{1}{c|}{1}          & \multicolumn{1}{c|}{0}          & \multicolumn{1}{c|}{1}          & \multicolumn{1}{c|}{0}          & \multicolumn{1}{c|}{0}          &       \diagbox[width=3.3em,height=1.1em,dir=SW]{}{}     \\ \cline{2-8} 
                                & \textbf{5}                     & \multicolumn{1}{c|}{1}          & \multicolumn{1}{c|}{0}          & \multicolumn{1}{c|}{1}          & \multicolumn{1}{c|}{1}          & \multicolumn{1}{c|}{0}          &     \diagbox[width=3.3em,height=1.1em,dir=SW]{}{}       \\ \cline{2-8} 
                                & \textbf{6}                     & \multicolumn{1}{c|}{1}          & \multicolumn{1}{c|}{1}          & \multicolumn{1}{c|}{1}          & \multicolumn{1}{c|}{0}          & \multicolumn{1}{c|}{0}          &      \diagbox[width=3.3em,height=1.1em,dir=SW]{}{}      \\ \cline{2-8} 
                                & \textbf{7}                     & \multicolumn{1}{c|}{1}          & \multicolumn{1}{c|}{1}          & \multicolumn{1}{c|}{1}          & \multicolumn{1}{c|}{1}          & \multicolumn{1}{c|}{0}          &     \diagbox[width=3.3em,height=1.1em,dir=SW]{}{}       \\ \hline
\multirow{8}{*}{\textbf{4}}     & \textbf{8}                     & \multicolumn{1}{c|}{1}          & \multicolumn{1}{c|}{0}          & \multicolumn{1}{c|}{1}          &  \multicolumn{1}{c|}{0}          & \multicolumn{1}{c|}{1}          & 0          \\ \cline{2-8} 
                                & \textbf{9}                     & \multicolumn{1}{c|}{1}          & \multicolumn{1}{c|}{0}          & \multicolumn{1}{c|}{1}          & \multicolumn{1}{c|}{0}          & \multicolumn{1}{c|}{1}          & 1          \\ \cline{2-8} 
                                & \textbf{10}                    & \multicolumn{1}{c|}{1}          & \multicolumn{1}{c|}{0}          & \multicolumn{1}{c|}{1}          & \multicolumn{1}{c|}{1}          & \multicolumn{1}{c|}{1}          & 0          \\ \cline{2-8} 
                                & \textbf{11}                    & \multicolumn{1}{c|}{1}          & \multicolumn{1}{c|}{0}          & \multicolumn{1}{c|}{1}          & \multicolumn{1}{c|}{1}          & \multicolumn{1}{c|}{1}          & 1          \\ \cline{2-8} 
                                & \textbf{12}                    & \multicolumn{1}{c|}{1}          & \multicolumn{1}{c|}{1}          & \multicolumn{1}{c|}{1}          & \multicolumn{1}{c|}{0}          & \multicolumn{1}{c|}{1}          & 0          \\ \cline{2-8} 
                                & \textbf{13}                    & \multicolumn{1}{c|}{1}          & \multicolumn{1}{c|}{1}          & \multicolumn{1}{c|}{1}          & \multicolumn{1}{c|}{0}          & \multicolumn{1}{c|}{1}          & 1          \\ \cline{2-8} 
                                & \textbf{14}                    & \multicolumn{1}{c|}{1}          & \multicolumn{1}{c|}{1}          & \multicolumn{1}{c|}{1}          & \multicolumn{1}{c|}{1}          & \multicolumn{1}{c|}{1}          & 0          \\ \cline{2-8} 
                                & \textbf{15}                    & \multicolumn{1}{c|}{1}          & \multicolumn{1}{c|}{1}          & \multicolumn{1}{c|}{1}          & \multicolumn{1}{c|}{1}          & \multicolumn{1}{c|}{1}          & 1         
                            \\ \hline \hline 
\end{tabular}
\label{binarization}}
\end{table}

The codeword of each TreeNet mode is shown in Table. \ref{binarization}. At each branch, one bit is needed to represent if the parent network is better than child networks, which is indicated by $p$. If a child network works better, another bit is needed to represent which child network is better, which is indicated by $c$. Therefore, 1, 3, 5, and 6 bits are needed for coding TreeNet modes at level $1$, $2$, $3$, and $4$, respectively. The Context-Based Adaptive Binary Arithmetic Coding (CABAC) \cite{b38} is used to encode the codewords and a context model is built for a corresponding bit.
\section{Experiment Results}
In this section, the experiment results of TreeNet are presented and analyzed in detail. Firstly, we introduce the experiment setting briefly. Secondly, the training dataset is presented. Thirdly, the overall performance of TreeNet integrated into VVC and HEVC is presented, respectively. After that, the usage rate, model size, and memory cost are provided. Finally, the data clustering-driven training is investigated.

\subsection{Experiment Setting}
To test the performance of TreeNet, it is implemented into VVC reference software VTM-4.0 and HEVC reference software HM-16.9. Libtorch is used to perform the feed-forward of networks. All experiments comply with the common test conditions specified in \cite{b39} and \cite{b40} with QP set as {22, 27, 32, 37}, and the all-intra main configuration is deployed. 24 video sequences with different resolutions, grouped as $Class A$, $B$, $C$, $D$, and $E$, are utilized for experiments, and the first frame of each sequence is tested. Bj$\phi$ntegaard distortion rate (BD-rate) \cite{b41} is used to evaluate the performance.

In our experiments, TreeNet is trained using the deep learning framework Pytorch \cite{b43} 1.8.0 on NVIDIA GeForce RTX 3060 GPU. The models for small blocks (4$\times$4 and 8$\times$8) are trained with a batch size of 512, while the models for large blocks (16$\times$16 to 64$\times$64) are trained with a batch size of 256.

\subsection{Training Dataset}
The training data is generated from pictures in the New York city library image set \cite{b42}. Firstly, these images are converted to YUV 4:2:0 color format. Then all training images are encoded with all intra configuration. The QP is set to 22, 27, 32, and 37. For all QPs, the encoder generates 5 different bitstreams for different block sizes from 4$\times$4 to 64$\times$64 separately, which will be used to train the models in TreeNet for different block sizes. The reconstructed reference samples of a block are extracted from the bitstreams on the decoder side as the network input. The original pixels of the block are taken as the ground truth.

It should be noted that, before the recursive training, all training data should be pre-processed by zero-centering to get rid of the low-frequency component and make the training easier. In \cite{b7} and \cite{b10}, blocks with complex textures are excluded from the training data since such data would degrade the training efficiency and practical performance, while in TreeNet training, all blocks are kept in the training data.

\begin{table*}[!htb]
\center
\caption{The BD-Rate Results of TreeNet over H.266/VVC}
\setlength{\tabcolsep}{4.2mm}
\renewcommand\arraystretch{1.15}
\scalebox{1}{
\begin{tabular}{cc|cccc|c|c}
\hline \hline
\multicolumn{2}{c|}{\multirow{2}{*}{\textbf{Sequence}}}                                                                             & \multicolumn{4}{c|}{\textbf{TreeNet}}                                                                                                            & \textbf{Fast TreeNet} & \textbf{MIP \cite{b12}} \\ \cline{3-8} 
\multicolumn{2}{c|}{}                                                                                                               & \multicolumn{1}{c|}{\textbf{\textit{Depth}=1}} & \multicolumn{1}{c|}{\textbf{\textit{Depth}=2}} & \multicolumn{1}{c|}{\textbf{\textit{Depth}=3}} & \textbf{\textit{Depth}=4} & \textbf{\textit{Depth}=3}    & -            \\ \hline
\multicolumn{1}{c|}{\multirow{6}{*}{\textbf{\begin{tabular}[c]{@{}c@{}}Class A\\ (4K)\end{tabular}}}}    & \textbf{Tango2}          & \multicolumn{1}{c|}{-4.4\%}             & \multicolumn{1}{c|}{-5.5\%}             & \multicolumn{1}{c|}{-6.2\%}             & -6.3\%             & -6.0\%                & -1.5\%       \\
\multicolumn{1}{c|}{}                                                                                    & \textbf{FoodMarket4}     & \multicolumn{1}{c|}{-6.2\%}             & \multicolumn{1}{c|}{-7.8\%}             & \multicolumn{1}{c|}{-8.3\%}             & -8.5\%             & -8.1\%                & -1.3\%       \\
\multicolumn{1}{c|}{}                                                                                    & \textbf{Campfire}        & \multicolumn{1}{c|}{-1.9\%}             & \multicolumn{1}{c|}{-2.4\%}             & \multicolumn{1}{c|}{-2.7\%}             & -2.8\%             & -2.6\%                & -0.7\%       \\
\multicolumn{1}{c|}{}                                                                                    & \textbf{CatRobot}        & \multicolumn{1}{c|}{-1.8\%}             & \multicolumn{1}{c|}{-2.6\%}             & \multicolumn{1}{c|}{-3.0\%}             & -3.1\%             & -2.9\%                & -0.7\%       \\
\multicolumn{1}{c|}{}                                                                                    & \textbf{DaylightRoad2}   & \multicolumn{1}{c|}{-2.0\%}             & \multicolumn{1}{c|}{-2.7\%}             & \multicolumn{1}{c|}{-3.0\%}             & -3.0\%             & -2.9\%                & -0.6\%       \\
\multicolumn{1}{c|}{}                                                                                    & \textbf{ParkRunning3}    & \multicolumn{1}{c|}{-1.3\%}             & \multicolumn{1}{c|}{-1.7\%}             & \multicolumn{1}{c|}{-2.2\%}             & -2.4\%             & -2.0\%                & -0.6\%       \\ \hline
\multicolumn{1}{c|}{\multirow{7}{*}{\textbf{\begin{tabular}[c]{@{}c@{}}Class B\\ (1080P)\end{tabular}}}} & \textbf{MarketPlace}     & \multicolumn{1}{c|}{-1.8\%}             & \multicolumn{1}{c|}{-2.3\%}             & \multicolumn{1}{c|}{-2.7\%}             & -2.7\%             & -2.7\%                & -0.7\%       \\
\multicolumn{1}{c|}{}                                                                                    & \textbf{RitualDance}     & \multicolumn{1}{c|}{-2.7\%}             & \multicolumn{1}{c|}{-3.3\%}             & \multicolumn{1}{c|}{-3.7\%}             & -3.8\%             & -3.6\%                & -1.0\%       \\
\multicolumn{1}{c|}{}                                                                                    & \textbf{Kimono}          & \multicolumn{1}{c|}{-1.6\%}             & \multicolumn{1}{c|}{-2.3\%}             & \multicolumn{1}{c|}{-2.6\%}             & -2.8\%             & -2.6\%                & -            \\
\multicolumn{1}{c|}{}                                                                                    & \textbf{ParkScene}       & \multicolumn{1}{c|}{-1.4\%}             & \multicolumn{1}{c|}{-2.0\%}             & \multicolumn{1}{c|}{-2.4\%}             & -2.4\%             & -2.3\%                & -            \\
\multicolumn{1}{c|}{}                                                                                    & \textbf{Cactus}          & \multicolumn{1}{c|}{-1.4\%}             & \multicolumn{1}{c|}{-2.0\%}             & \multicolumn{1}{c|}{-2.5\%}             & -2.6\%             & -2.3\%                & -0.7\%       \\
\multicolumn{1}{c|}{}                                                                                    & \textbf{BasketballDrive} & \multicolumn{1}{c|}{-1.9\%}             & \multicolumn{1}{c|}{-2.2\%}             & \multicolumn{1}{c|}{-2.5\%}             & -2.6\%             & -2.4\%                & -0.6\%       \\
\multicolumn{1}{c|}{}                                                                                    & \textbf{BQTerrace}       & \multicolumn{1}{c|}{-1.2\%}             & \multicolumn{1}{c|}{-1.6\%}             & \multicolumn{1}{c|}{-2.0\%}             & -1.9\%             & -2.0\%                & -0.4\%       \\ \hline
\multicolumn{1}{c|}{\multirow{4}{*}{\textbf{\begin{tabular}[c]{@{}c@{}}Class C\\ (WVGA)\end{tabular}}}}  & \textbf{BasketballDrill} & \multicolumn{1}{c|}{-1.0\%}             & \multicolumn{1}{c|}{-1.3\%}             & \multicolumn{1}{c|}{-1.7\%}             & -1.6\%             & -1.6\%                & -0.7\%       \\
\multicolumn{1}{c|}{}                                                                                    & \textbf{BQMall}          & \multicolumn{1}{c|}{-1.4\%}             & \multicolumn{1}{c|}{-1.8\%}             & \multicolumn{1}{c|}{-2.1\%}             & -2.2\%             & -2.0\%                & -0.8\%       \\
\multicolumn{1}{c|}{}                                                                                    & \textbf{PartyScene}      & \multicolumn{1}{c|}{-1.3\%}             & \multicolumn{1}{c|}{-1.7\%}             & \multicolumn{1}{c|}{-2.0\%}             & -2.0\%             & -1.9\%                & -0.7\%       \\
\multicolumn{1}{c|}{}                                                                                    & \textbf{RaceHorses}      & \multicolumn{1}{c|}{-1.6\%}             & \multicolumn{1}{c|}{-2.0\%}             & \multicolumn{1}{c|}{-2.2\%}             & -2.3\%             & -2.2\%                & -0.7\%       \\ \hline
\multicolumn{1}{c|}{\multirow{4}{*}{\textbf{\begin{tabular}[c]{@{}c@{}}Class D\\ (WQVGA)\end{tabular}}}} & \textbf{BasketballPass}  & \multicolumn{1}{c|}{-1.2\%}             & \multicolumn{1}{c|}{-1.3\%}             & \multicolumn{1}{c|}{-1.6\%}             & -1.6\%             & -1.5\%                & -0.8\%       \\
\multicolumn{1}{c|}{}                                                                                    & \textbf{BQSquare}        & \multicolumn{1}{c|}{-0.8\%}             & \multicolumn{1}{c|}{-1.2\%}             & \multicolumn{1}{c|}{-1.5\%}             & -1.5\%             & -1.5\%                & -0.8\%       \\
\multicolumn{1}{c|}{}                                                                                    & \textbf{BlowingBubbles}  & \multicolumn{1}{c|}{-1.6\%}             & \multicolumn{1}{c|}{-1.7\%}             & \multicolumn{1}{c|}{-1.9\%}             & -1.9\%             & -1.7\%                & -0.7\%       \\
\multicolumn{1}{c|}{}                                                                                    & \textbf{RaceHorses}      & \multicolumn{1}{c|}{-1.9\%}             & \multicolumn{1}{c|}{-2.4\%}             & \multicolumn{1}{c|}{-2.7\%}             & -2.7\%             & -2.6\%                & -0.9\%       \\ \hline
\multicolumn{1}{c|}{\multirow{3}{*}{\textbf{\begin{tabular}[c]{@{}c@{}}Class E\\ (720P)\end{tabular}}}}  & \textbf{FourPeople}      & \multicolumn{1}{c|}{-2.0\%}             & \multicolumn{1}{c|}{-2.6\%}             & \multicolumn{1}{c|}{-3.0\%}             & -3.0\%             & -2.9\%                & -0.8\%       \\
\multicolumn{1}{c|}{}                                                                                    & \textbf{Johnny}          & \multicolumn{1}{c|}{-2.4\%}             & \multicolumn{1}{c|}{-2.7\%}             & \multicolumn{1}{c|}{-3.2\%}             & -3.0\%             & -3.0\%                & -0.9\%       \\
\multicolumn{1}{c|}{}                                                                                    & \textbf{KristenAndSara}  & \multicolumn{1}{c|}{-1.6\%}             & \multicolumn{1}{c|}{-2.2\%}             & \multicolumn{1}{c|}{-2.7\%}             & -2.7\%             & -2.6\%                & -0.8\%       \\ \hline
\multicolumn{2}{c|}{\textbf{Average}}                                                                                               & \multicolumn{1}{c|}{-1.9\%}             & \multicolumn{1}{c|}{-2.5\%}             & \multicolumn{1}{c|}{-2.9\%}             & -2.9\%             & -2.8\%                & -0.8\%      
   \\ \hline \hline  
\end{tabular}  
}
\label{table2}
\end{table*}

\subsection{Comparison with H.266/VVC}
In the experiments, we first test the performance of applying TreeNet with different depths to the VTM-4.0. Then the performance of the fast termination strategy is provided.

The experimental results are shown in Table \ref{table2}. As observed, TreeNet with $Depth$=4, 3, 2, and 1 can bring an average BD-rate saving of 2.9\%, 2.9\%, 2.5\%, and 1.9\% for the luma component, respectively. For all test sequences, TreeNet with $Depth$=4 brings similar gains as TreeNet with $Depth$=3. Therefore, TreeNet with $Depth$=3 is adopted and used in the following experiments. For simplicity, we call TreeNet with $Depth$=3 as TreeNet in the following. The performance of TreeNet with the fast termination strategy (fast TreeNet) is degraded to an average of 2.8\% BD-rate saving (up to 8.1\%), however the encoding time is much reduced, which is shown in Table \ref{table4}. Compared with MIP \cite{b12}, which has simpler network architecture and more network modes than TreeNet, the BD-rate improvement brought by TreeNet is significant, reaching 2.0\% improvement on average.

\begin{table}[!t]
\center
\caption{Running Time Comparison Over VVC}
\renewcommand\arraystretch{1.15}
\setlength{\tabcolsep}{2.5mm}
\scalebox{0.95}{
\begin{threeparttable}
\begin{tabular}{cc|c|c|c}
\hline \hline
\multicolumn{2}{c|}{\textbf{Anchor}}                                                   & \textbf{\textit{Depth}} & \textbf{Encoding Time} & \textbf{Decoding Time} \\ \hline
\multicolumn{1}{c|}{\multirow{6}{*}{\textbf{VTM}}} & \multirow{4}{*}{\textbf{TreeNet}} & 1              & 753\%                  & 3909\%                 \\ \cline{3-5} 
\multicolumn{1}{c|}{}                              &                                   & 2              & 1872\%                 & 4613\%                 \\
\cline{3-5} 
\multicolumn{1}{c|}{}                              &                                   & 3              & 4115\%                 & 5173\%                 \\\cline{3-5} 
\multicolumn{1}{c|}{}                              &                                   & 4              & 8632\%                 & 5294\%                 \\ \cline{2-5} 
\multicolumn{1}{c|}{}                              & \textbf{Fast TreeNet}             & 3              & 2493\%                 & 5061\%                               \\
\cline{2-5} 
\multicolumn{1}{c|}{}                              & \textbf{MIP \cite{b12}}             & -              & 138\%                 & 99\%                               \\
\hline \hline                 
\end{tabular}
\begin{tablenotes}
\footnotesize
\item[*]Anchor is 100\%
\end{tablenotes}
\end{threeparttable}}
\label{table4}
\end{table}

The running times of these tests are summarized in Table \ref{table4}. All tests for TreeNet are done with an AMD Ryzen7 5800H CPU and an NVIDIA GeForce RTX 3060 GPU. MIP is conducted on an Intel Xeon cluster (E5-2697A v4). As TreeNet goes deeper, the encoding time increases exponentially. The encoding time of TreeNet with $Depth$=3 is around 41 times of the anchor. With the fast termination strategy, the encoding time is reduced to around 25 times. The decoding time of fast TreeNet is roughly 50 times of the anchor. Compared with MIP, both the encoding and decoding time increase greatly due to the complex network structure in TreeNet. How to simplify the network structure in TreeNet is our future work. 

\begin{figure}[t!]
\centerline{\includegraphics[width=9.3cm, height=4.8cm]{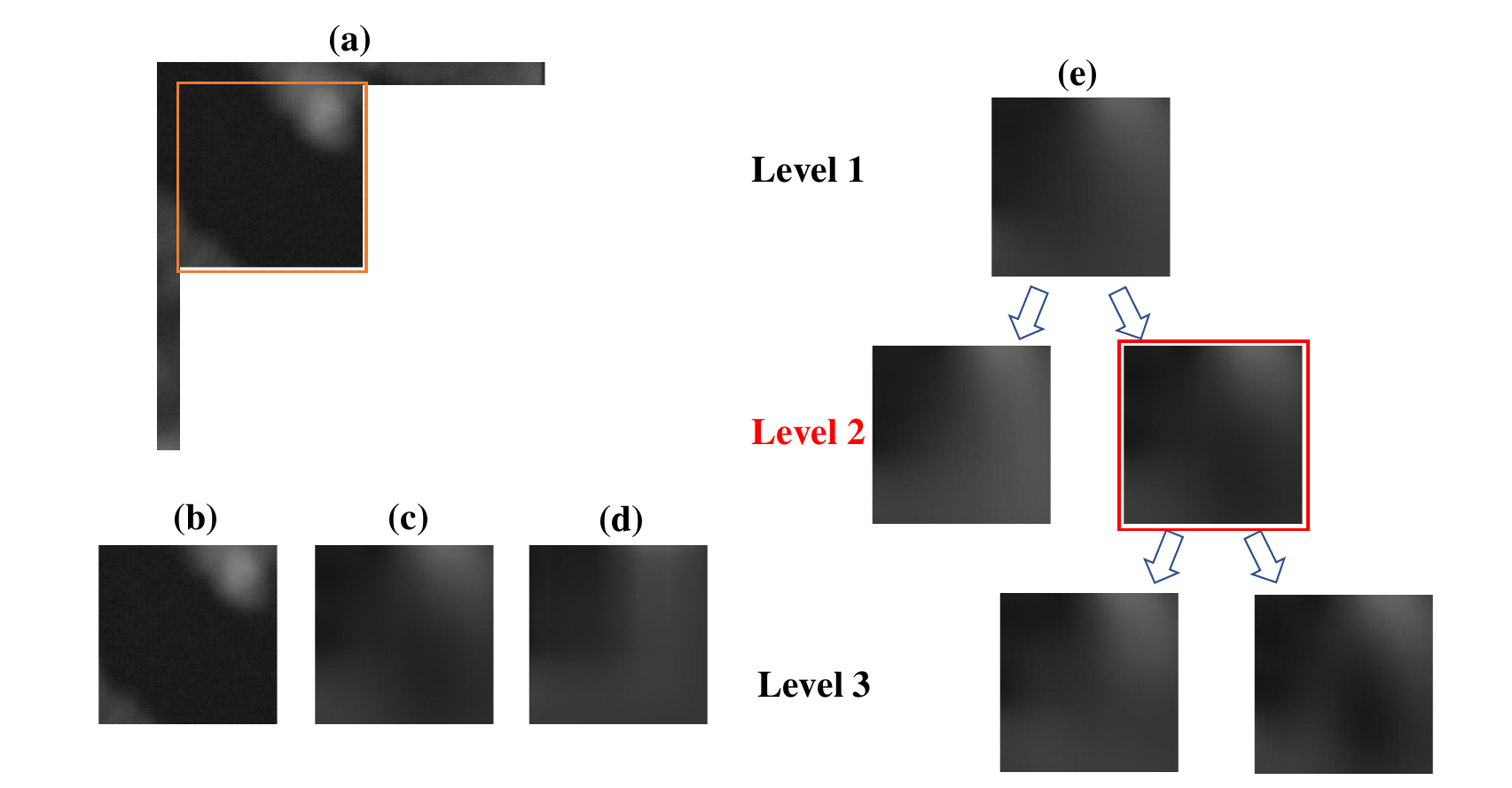}}
\caption{Prediction of a 64$\times$64 block (a) 8-line reference samples and original pixels. (b) original pixels (c) prediction via the best TreeNet mode at level $2$ (MSE: 198.4) (d) prediction via the best VVC mode of index 0 (MSE: 518.7) (e) predictions via the TreeNet modes at level $1$ (MSE: 472.9), level $2$ (MSE: 1045.5, 198.4), and level $3$ (MSE: 519.7, 194.0)}
\label{64}
\end{figure}

\begin{figure}[!htb]
\centerline{\includegraphics[width=9.3cm, height=4.8cm]{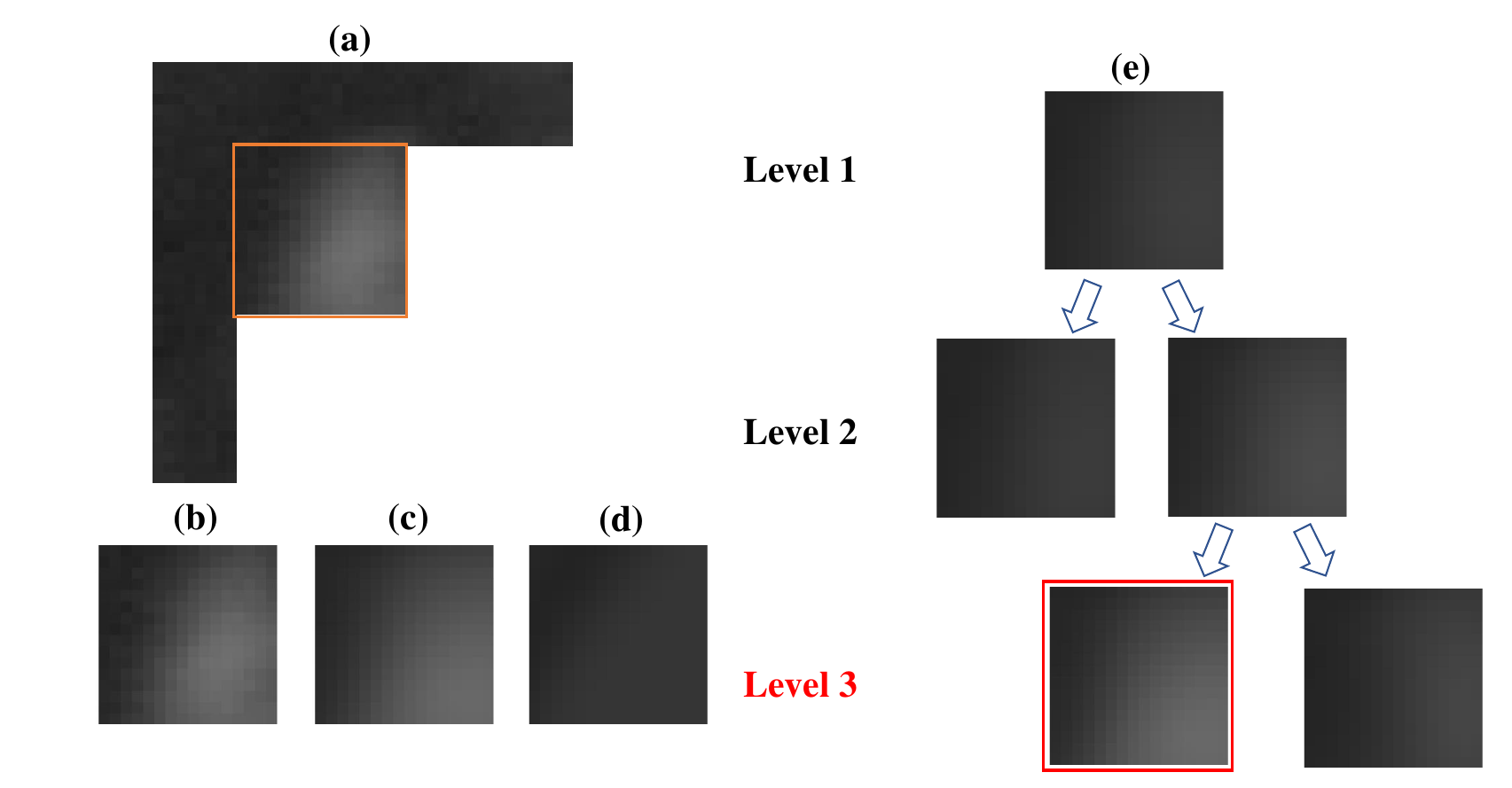}}
\caption{Prediction of a 16$\times$16 block (a) 8-line reference samples and original pixels. (b) original pixels (c) prediction via the best TreeNet mode at level $3$ (MSE: 53.7) (d) prediction via the best VVC mode of index 63 (MSE: 905.7) (e) predictions via the TreeNet modes at level $1$ (MSE: 721.4), level $2$ (MSE: 815.6, 401.0), and level $3$ (MSE: 53.7, 621.4)}
\label{16}
\end{figure}

\begin{figure}[!htb]
\centerline{\includegraphics[width=9.3cm, height=4.8cm]{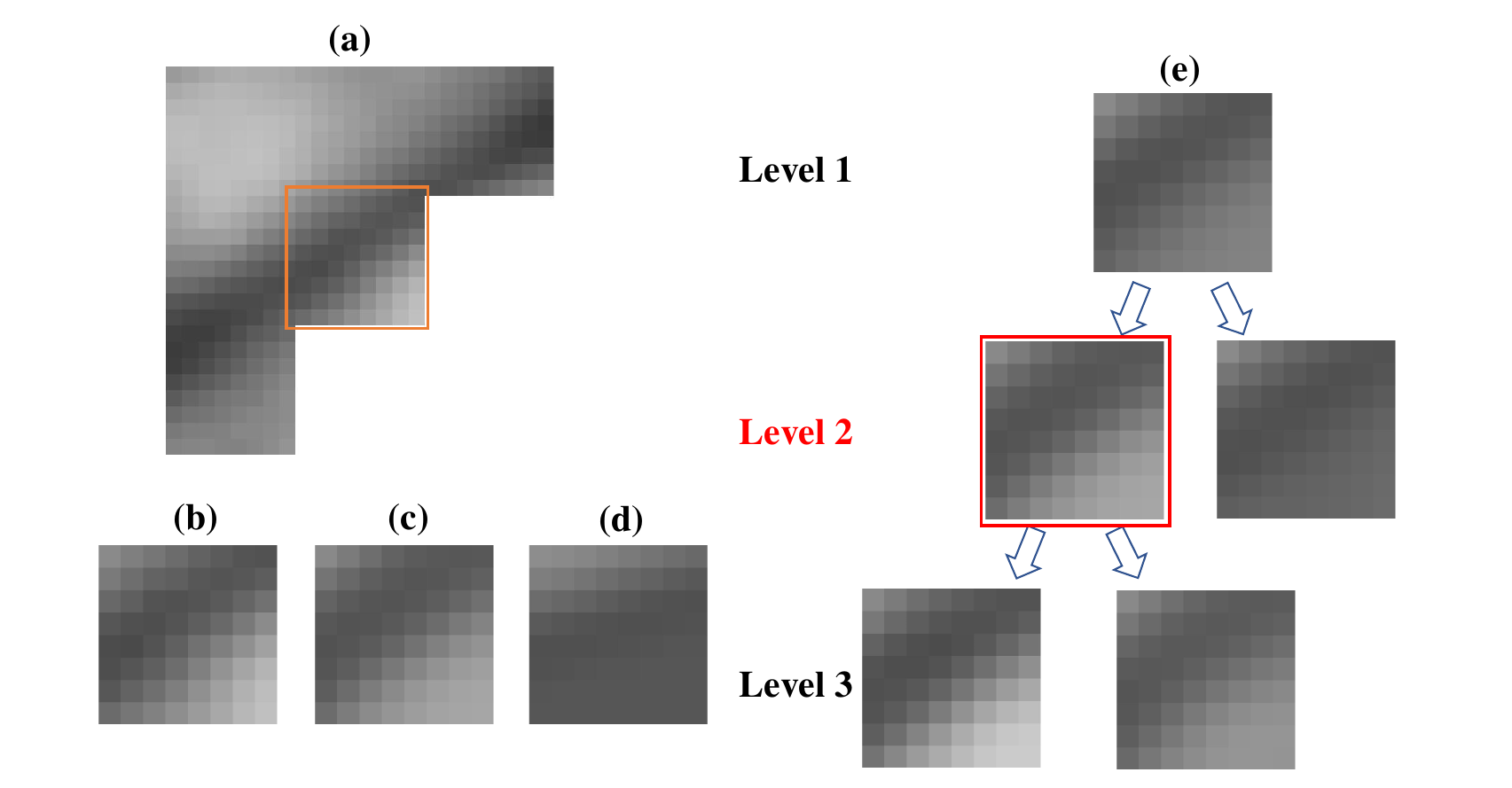}}
\caption{Prediction of a 8$\times$8 block (a) 8-line reference samples and original pixels. (b) original pixels (c) prediction via the best TreeNet mode at level $2$ (MSE: 71.1) (d) prediction via the best VVC mode of index 11 (MSE: 518.7) (e) predictions via the TreeNet modes at level $1$ (MSE: 415.4), level $2$ (MSE: 71.1, 1069.3), and level $3$ (MSE: 160.0, 172.7)}
\label{8}
\end{figure}

In addition to the BD-rate saving, we also provide some visual examples of prediction results of TreeNet and compare them with VVC intra prediction. The examples are taken from the first frame of \textit{Kimono} encoded with QP=27. The comparison results are shown in Figs. \ref{64}, \ref{16}, and \ref{8}. As observed, TreeNet could produce more accurate predictions than the VVC directional prediction modes. As shown in Fig. \ref{64}, the prediction of TreeNet distinguishes the black and white areas more accurately and clearly than the directional prediction of VVC. In Fig. \ref{16}, TreeNet generates the predictions that are closer to the original textures compared with the VVC intra prediction. In Fig. \ref{8}, there is an obvious directional pattern in the original block. TreeNet could still generate better predictions than the VVC directional prediction mode.

\begin{table*}[htb]
\center
\caption{The BD-Rate Results of Fast TreeNet and Other Approaches over H.265/HEVC}
\setlength{\tabcolsep}{2.0mm}
\renewcommand\arraystretch{1.15}
\scalebox{1}{
\begin{tabular}{cc|ccc|cccc}
\hline
\hline
\multicolumn{2}{c|}{\multirow{2}{*}{\textbf{Sequence}}}                                                                             & \multicolumn{3}{c|}{\textbf{HM}}                                                                                  & \multicolumn{4}{c}{\textbf{HM with 8x8 constraint}}                                                                                                             \\ \cline{3-9} 
\multicolumn{2}{c|}{}                                                                                                               & \multicolumn{1}{c|}{\textbf{Fast TreeNet}} & \multicolumn{1}{c|}{\textbf{IPFCN-D\cite{b7}}} & \textbf{PS-RNN\cite{PSRNN}} & \multicolumn{1}{c|}{\textbf{Fast TreeNet}} & \multicolumn{1}{c|}{\textbf{NM\cite{TMM}}} & \multicolumn{1}{c|}{\textbf{IPFCN-D\cite{b7}}} & \textbf{PS-RNN\cite{PSRNN}} \\ \hline
\multicolumn{1}{c|}{\multirow{8}{*}{\textbf{\begin{tabular}[c]{@{}c@{}}Class A\\ (4K)\end{tabular}}}}    & \textbf{Tango}           & \multicolumn{1}{c|}{-8.2\%}                & \multicolumn{1}{c|}{-7.4\%}                & -                       & \multicolumn{1}{c|}{-20.1\%}               & \multicolumn{1}{c|}{-18.4\%}             & \multicolumn{1}{c|}{-15.6\%}               & -                       \\
\multicolumn{1}{c|}{}                                                                                    & \textbf{Drums}           & \multicolumn{1}{c|}{-4.4\%}                & \multicolumn{1}{c|}{-3.8\%}                & -                       & \multicolumn{1}{c|}{-8.6\%}                & \multicolumn{1}{c|}{-6.7\%}              & \multicolumn{1}{c|}{-2.5\%}                & -                       \\
\multicolumn{1}{c|}{}                                                                                    & \textbf{CampfireParty}   & \multicolumn{1}{c|}{-4.3\%}                & \multicolumn{1}{c|}{-3.0\%}                & -                       & \multicolumn{1}{c|}{-6.4\%}                & \multicolumn{1}{c|}{-7.7\%}              & \multicolumn{1}{c|}{-4.3\%}                & -                       \\
\multicolumn{1}{c|}{}                                                                                    & \textbf{ToddlerFountain} & \multicolumn{1}{c|}{-4.4\%}                & \multicolumn{1}{c|}{-3.4\%}                & -                       & \multicolumn{1}{c|}{-5.8\%}                & \multicolumn{1}{c|}{-5.7\%}              & \multicolumn{1}{c|}{-3.1\%}                & -                       \\
\multicolumn{1}{c|}{}                                                                                    & \textbf{CatRobot}        & \multicolumn{1}{c|}{-6.0\%}                & \multicolumn{1}{c|}{-4.2\%}                & -                       & \multicolumn{1}{c|}{-9.7\%}                & \multicolumn{1}{c|}{-8.0\%}              & \multicolumn{1}{c|}{-5.7\%}                & -                       \\
\multicolumn{1}{c|}{}                                                                                    & \textbf{TrafficFlow}     & \multicolumn{1}{c|}{-6.5\%}                & \multicolumn{1}{c|}{-4.2\%}                & -                       & \multicolumn{1}{c|}{-12.7\%}               & \multicolumn{1}{c|}{-8.4\%}              & \multicolumn{1}{c|}{-7.2\%}                & -                       \\
\multicolumn{1}{c|}{}                                                                                    & \textbf{DaylightRoad}    & \multicolumn{1}{c|}{-6.4\%}                & \multicolumn{1}{c|}{-4.5\%}                & -                       & \multicolumn{1}{c|}{-8.4\%}                & \multicolumn{1}{c|}{-8.5\%}              & \multicolumn{1}{c|}{-7.0\%}                & -                       \\
\multicolumn{1}{c|}{}                                                                                    & \textbf{Rollercoaster}   & \multicolumn{1}{c|}{-5.9\%}                & \multicolumn{1}{c|}{-5.8\%}                & -                       & \multicolumn{1}{c|}{-18.9\%}               & \multicolumn{1}{c|}{-17.8\%}             & \multicolumn{1}{c|}{-11.6\%}               & -                       \\ \hline
\multicolumn{1}{c|}{\multirow{5}{*}{\textbf{\begin{tabular}[c]{@{}c@{}}Class B\\ 1080P\end{tabular}}}}   & \textbf{Kimono}          & \multicolumn{1}{c|}{-4.6\%}                & \multicolumn{1}{c|}{-3.1\%}                & -2.2\%                  & \multicolumn{1}{c|}{-14.9\%}               & \multicolumn{1}{c|}{-10.9\%}             & \multicolumn{1}{c|}{-5.5\%}                & -6.6\%                  \\
\multicolumn{1}{c|}{}                                                                                    & \textbf{ParkScene}       & \multicolumn{1}{c|}{-4.8\%}                & \multicolumn{1}{c|}{-3.6\%}                & -2.8\%                  & \multicolumn{1}{c|}{-4.9\%}                & \multicolumn{1}{c|}{-4.4\%}              & \multicolumn{1}{c|}{-2.3\%}                & -3.4\%                  \\
\multicolumn{1}{c|}{}                                                                                    & \textbf{Cactus}          & \multicolumn{1}{c|}{-4.6\%}                & \multicolumn{1}{c|}{-3.2\%}                & -2.5\%                  & \multicolumn{1}{c|}{-5.8\%}                & \multicolumn{1}{c|}{-4.3\%}              & \multicolumn{1}{c|}{-2.0\%}                & -3.3\%                  \\
\multicolumn{1}{c|}{}                                                                                    & \textbf{BasketballDrive} & \multicolumn{1}{c|}{-5.4\%}                & \multicolumn{1}{c|}{-3.6\%}                & -1.8\%                  & \multicolumn{1}{c|}{-11.6\%}               & \multicolumn{1}{c|}{-9.9\%}              & \multicolumn{1}{c|}{-4.6\%}                & -7.8\%                  \\
\multicolumn{1}{c|}{}                                                                                    & \textbf{BQTerrace}       & \multicolumn{1}{c|}{-3.8\%}                & \multicolumn{1}{c|}{-2.1\%}                & -2.6\%                  & \multicolumn{1}{c|}{-3.7\%}                & \multicolumn{1}{c|}{-3.2\%}              & \multicolumn{1}{c|}{-1.6\%}                & -2.6\%                  \\ \hline
\multicolumn{1}{c|}{\multirow{4}{*}{\textbf{\begin{tabular}[c]{@{}c@{}}Class C\\ (WVGA)\end{tabular}}}}  & \textbf{BasketballDrill} & \multicolumn{1}{c|}{-2.9\%}                & \multicolumn{1}{c|}{-1.5\%}                & -2.1\%                  & \multicolumn{1}{c|}{-2.9\%}                & \multicolumn{1}{c|}{-1.6\%}              & \multicolumn{1}{c|}{1.2\%}                 & -2.9\%                  \\
\multicolumn{1}{c|}{}                                                                                    & \textbf{BQMall}          & \multicolumn{1}{c|}{-4.0\%}                & \multicolumn{1}{c|}{-2.2\%}                & -3.1\%                  & \multicolumn{1}{c|}{-3.6\%}                & \multicolumn{1}{c|}{-3.5\%}              & \multicolumn{1}{c|}{-1.4\%}                & -2.9\%                  \\
\multicolumn{1}{c|}{}                                                                                    & \textbf{PartyScene}      & \multicolumn{1}{c|}{-3.0\%}                & \multicolumn{1}{c|}{-1.6\%}                & -2.6\%                  & \multicolumn{1}{c|}{-2.9\%}                & \multicolumn{1}{c|}{-2.4\%}              & \multicolumn{1}{c|}{-1.1\%}                & -2.3\%                  \\
\multicolumn{1}{c|}{}                                                                                    & \textbf{RaceHorsesC}     & \multicolumn{1}{c|}{-3.5\%}                & \multicolumn{1}{c|}{-3.2\%}                & -1.4\%                  & \multicolumn{1}{c|}{-4.1\%}                & \multicolumn{1}{c|}{-3.1\%}              & \multicolumn{1}{c|}{-1.2\%}                & -2.8\%                  \\ \hline
\multicolumn{1}{c|}{\multirow{4}{*}{\textbf{\begin{tabular}[c]{@{}c@{}}Class D\\ (WQVGA)\end{tabular}}}} & \textbf{BasketballPass}  & \multicolumn{1}{c|}{-3.6\%}                & \multicolumn{1}{c|}{-1.2\%}                & -1.8\%                  & \multicolumn{1}{c|}{-3.0\%}                & \multicolumn{1}{c|}{-2.7\%}              & \multicolumn{1}{c|}{0.4\%}                 & -2.5\%                  \\
\multicolumn{1}{c|}{}                                                                                    & \textbf{BQSquare}        & \multicolumn{1}{c|}{-2.7\%}                & \multicolumn{1}{c|}{-0.9\%}                & -2.0\%                  & \multicolumn{1}{c|}{-1.5\%}                & \multicolumn{1}{c|}{-1.6\%}              & \multicolumn{1}{c|}{-0.8\%}                & -1.8\%                  \\
\multicolumn{1}{c|}{}                                                                                    & \textbf{BlowingBubbles}  & \multicolumn{1}{c|}{-3.2\%}                & \multicolumn{1}{c|}{-1.9\%}                & -2.8\%                  & \multicolumn{1}{c|}{-3.1\%}                & \multicolumn{1}{c|}{-2.7\%}              & \multicolumn{1}{c|}{-1.3\%}                & -2.3\%                  \\
\multicolumn{1}{c|}{}                                                                                    & \textbf{RaceHorses}      & \multicolumn{1}{c|}{-4.8\%}                & \multicolumn{1}{c|}{-3.2\%}                & -3.5\%                  & \multicolumn{1}{c|}{-4.0\%}                & \multicolumn{1}{c|}{-3.1\%}              & \multicolumn{1}{c|}{-1.0\%}                & -2.6\%                  \\ \hline
\multicolumn{1}{c|}{\multirow{3}{*}{\textbf{\begin{tabular}[c]{@{}c@{}}Class E\\ (720P)\end{tabular}}}}  & \textbf{FourPeople}      & \multicolumn{1}{c|}{-6.6\%}                & \multicolumn{1}{c|}{-4.4\%}                & -3.8\%                  & \multicolumn{1}{c|}{-6.9\%}                & \multicolumn{1}{c|}{-6.0\%}              & \multicolumn{1}{c|}{-3.4\%}                & -6.8\%                  \\
\multicolumn{1}{c|}{}                                                                                    & \textbf{Johnny}          & \multicolumn{1}{c|}{-7.2\%}                & \multicolumn{1}{c|}{-5.3\%}                & -4.0\%                  & \multicolumn{1}{c|}{-9.5\%}                & \multicolumn{1}{c|}{-8.6\%}              & \multicolumn{1}{c|}{-8.3\%}                & -5.6\%                  \\
\multicolumn{1}{c|}{}                                                                                    & \textbf{KristenAndSara}  & \multicolumn{1}{c|}{-5.8\%}                & \multicolumn{1}{c|}{-3.9\%}                & -3.2\%                  & \multicolumn{1}{c|}{-7.6\%}                & \multicolumn{1}{c|}{-6.7\%}              & \multicolumn{1}{c|}{-5.8\%}                & -6.6\%                  \\ \hline
\multicolumn{2}{c|}{\textbf{Average}}                                                                                               & \multicolumn{1}{c|}{-4.9\%}                & \multicolumn{1}{c|}{-3.4\%}                & -2.7\%                  & \multicolumn{1}{c|}{-7.5\%}                & \multicolumn{1}{c|}{-6.5\%}              & \multicolumn{1}{c|}{-4.0\%}                & -3.9\%                \\\hline
\hline  
\end{tabular}
}
\label{table3}
\end{table*}

\subsection{Comparison with H.265/HEVC and The State of The Art}
We compare fast TreeNet with Intra Prediction Fully-Connected Networks (IPFCN) \cite{b7}, Progressive Spatial Recurrent Neural Network (PS-RNN) \cite{PSRNN}, and multiple neural network model-based enhanced intra prediction (NM) \cite{TMM}. Among these approaches, NM in \cite{TMM} only allows 8$\times$8 intra coding, and such a manner is also applied in PS-RNN. In \cite{TMM}, IPFCN-D integrated into HM with 8$\times$8 constraint is also reimplemented. To make a more detailed comparison, fast TreeNet is implemented with and without 8$\times$8 constraint, respectively. The comparison results are presented in Table \ref{table3}.

As shown in Table \ref{table3}, for the luma component, fast TreeNet brings an average BD-rate saving of 4.9\% (up to 8.2\%). Compared with IPFCN-D, fast TreeNet achieves 1.5\% BD-rate reduction on average. Fast TreeNet can also bring an average of 1.7\% BD-rate reduction over PS-RNN on Class B, C, D, and E. When only 8$\times$8 intra coding is allowed, the improvement of fast TreeNet is more obvious, reaching an average of 7.5\% BD-rate reduction. Compared with IPFCN-D and NM, fast TreeNet achieves 3.5\% and 1.0\% BD-rate reduction on average, respectively. Fast TreeNet can also bring 1.7\% BD-rate reduction on average over PS-RNN on Class B, C, D, and E.

Furthermore, we replace all the directional prediction modes in HEVC with fast TreeNet and test the performance on the first frame of some high-resolution sequences. The results are presented in Table \ref{table6}. As shown in Table \ref{table6}, fast TreeNet with $Depth$=4 and 3 brings an average of 2.3\% and 2.1\% BD-rate reduction respectively, which indicates fast TreeNet replacing all the directional prediction modes still performs well for high-resolution sequences. The result also indicates that more networks should be adopted when TreeNet is used to replace the directional prediction modes than combining with them.

\begin{table}[!t]
\center
\caption{The BD-Rate Results of The Substitution Scheme}
\renewcommand\arraystretch{1.15}
\setlength{\tabcolsep}{4.8mm}
\scalebox{0.95}{
\begin{tabular}{cc|cc}
\hline \hline
\multicolumn{2}{c|}{\multirow{2}{*}{\textbf{Sequence}}}                                                                             & \multicolumn{2}{c}{\textbf{Fast TreeNet}}                    \\ \cline{3-4} 
\multicolumn{2}{c|}{}                                                                                                               & \multicolumn{1}{c|}{\textbf{\textit{Depth}=3}} & \textbf{\textit{Depth}=4} \\ \hline
\multicolumn{1}{c|}{\multirow{10}{*}{\textbf{\begin{tabular}[c]{@{}c@{}}Class A\\ (4K)\end{tabular}}}}   & \textbf{Tango}           & \multicolumn{1}{c|}{-4.8\%}            & -5.0\%            \\
\multicolumn{1}{c|}{}                                                                                    & \textbf{Drums}           & \multicolumn{1}{c|}{-1.6\%}            & -1.9\%            \\
\multicolumn{1}{c|}{}                                                                                    & \textbf{CampfireParty}   & \multicolumn{1}{c|}{-1.2\%}            & -1.4\%            \\
\multicolumn{1}{c|}{}                                                                                    & \textbf{ToddlerFountain} & \multicolumn{1}{c|}{-3.0\%}            & -3.2\%            \\
\multicolumn{1}{c|}{}                                                                                    & \textbf{CatRobot}        & \multicolumn{1}{c|}{-1.4\%}            & -1.6\%            \\
\multicolumn{1}{c|}{}                                                                                    & \textbf{TrafficFlow}     & \multicolumn{1}{c|}{-1.5\%}            & -1.6\%            \\
\multicolumn{1}{c|}{}                                                                                    & \textbf{FaylightRoad}    & \multicolumn{1}{c|}{-2.7\%}            & -3.1\%            \\
\multicolumn{1}{c|}{}                                                                                    & \textbf{Rollercoaster}   & \multicolumn{1}{c|}{-1.9\%}            & -2.2\%            \\
\multicolumn{1}{c|}{}                                                                                    & \textbf{Traffic}         & \multicolumn{1}{c|}{-2.9\%}            & -3.1\%            \\
\multicolumn{1}{c|}{}                                                                                    & \textbf{PeopleOnStreet}  & \multicolumn{1}{c|}{-3.4\%}            & -3.4\%            \\ \hline
\multicolumn{1}{c|}{\multirow{5}{*}{\textbf{\begin{tabular}[c]{@{}c@{}}Class B\\ (1080P)\end{tabular}}}} & \textbf{Kimono}          & \multicolumn{1}{c|}{-2.5\%}            & -2.8\%            \\
\multicolumn{1}{c|}{}                                                                                    & \textbf{ParkScene}       & \multicolumn{1}{c|}{-2.6\%}            & -2.9\%            \\
\multicolumn{1}{c|}{}                                                                                    & \textbf{Cactus}          & \multicolumn{1}{c|}{-0.8\%}            & -1.2\%            \\
\multicolumn{1}{c|}{}                                                                                    & \textbf{BasketballDrive} & \multicolumn{1}{c|}{-0.9\%}            & -1.1\%            \\
\multicolumn{1}{c|}{}                                                                                    & \textbf{BQTerrace}       & \multicolumn{1}{c|}{-0.2\%}            & -0.3\%            \\ \hline
\multicolumn{2}{c|}{\textbf{Average}}                                                                                               & \multicolumn{1}{c|}{-2.1\%}            & -2.3\%      \\\hline \hline     
\end{tabular}}
\label{table6}
\end{table}

\subsection{Usage Rate}
We calculate the usage rate of fast TreeNet at the pixel level when combining with the VVC directional prediction modes, which can be calculated by:
\begin{small}
\begin{equation}
\eta =\frac{\sum_{i=1}^{14}( n_{i}\times w_{i} \times h_{i})}{W\times H} \times 100\%
\end{equation}
\end{small}where $\eta $ denotes the usage rate. $W$ and $H$ denote the width and height of each frame. $n_{i}$ represents the number of blocks with a size of $w_{i}\times h_{i}$ coded by fast TreeNet in each frame. Since MTT \cite{b26} is enabled in VTM, there are 14 blocks sizes supported. The QPs are set as $\left \{22, 27, 32, 37  \right \}$. As shown in Table \ref{table8}, the usage rates of fast TreeNet in all sequences are remarkable, especially in Class A. The average usage rate of fast TreeNet in all test sequences is 51.3\% (up to 71.7\%), which demonstrates its efficiency. 

\begin{figure*}[htbp]
\centerline{\includegraphics[width=18.3cm, height=9cm]{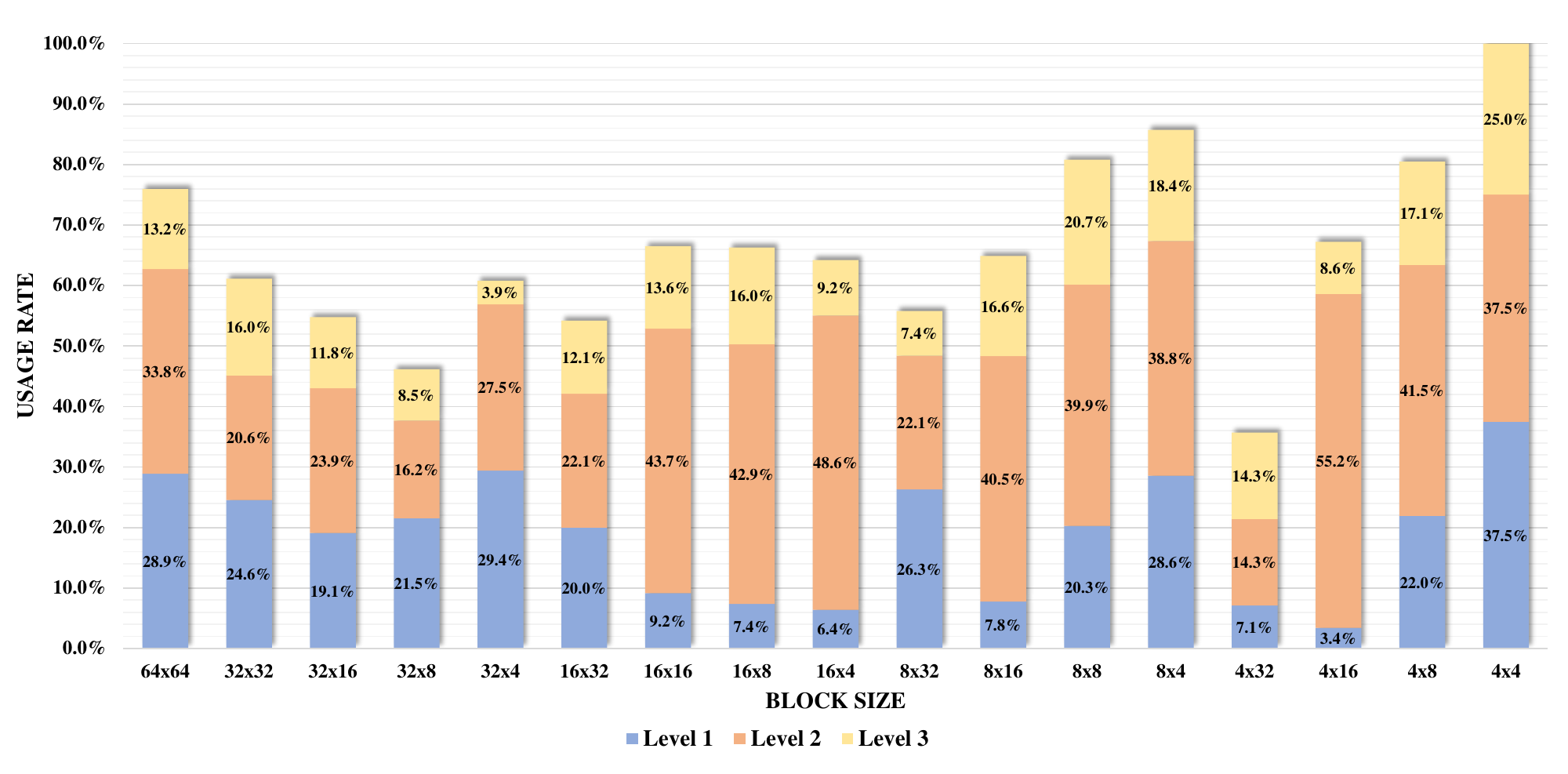}}
\caption{The usage rates of fast TreeNet in blocks of different sizes and the usage rate of networks at different levels on the first frame of \textit{Kimono} encoded with QP=32.}
\label{fig9}
\end{figure*}

Furthermore, we provide an example of the usage rates of fast TreeNet at different levels for different block sizes, which is shown in Fig. \ref{fig9}. We can observe that fast TreeNet has a significant usage rate in predicting blocks of all sizes. For blocks of size 4$\times$4, 4$\times$8, 8$\times$4, and 8$\times$8, the usage rates of fast TreeNet are more than 80\%, which demonstrates that fast TreeNet has more advantage in predicting small blocks than the directional prediction modes. 

\begin{figure*}[htbp]
\centerline{\includegraphics[width=19cm,height=11cm]{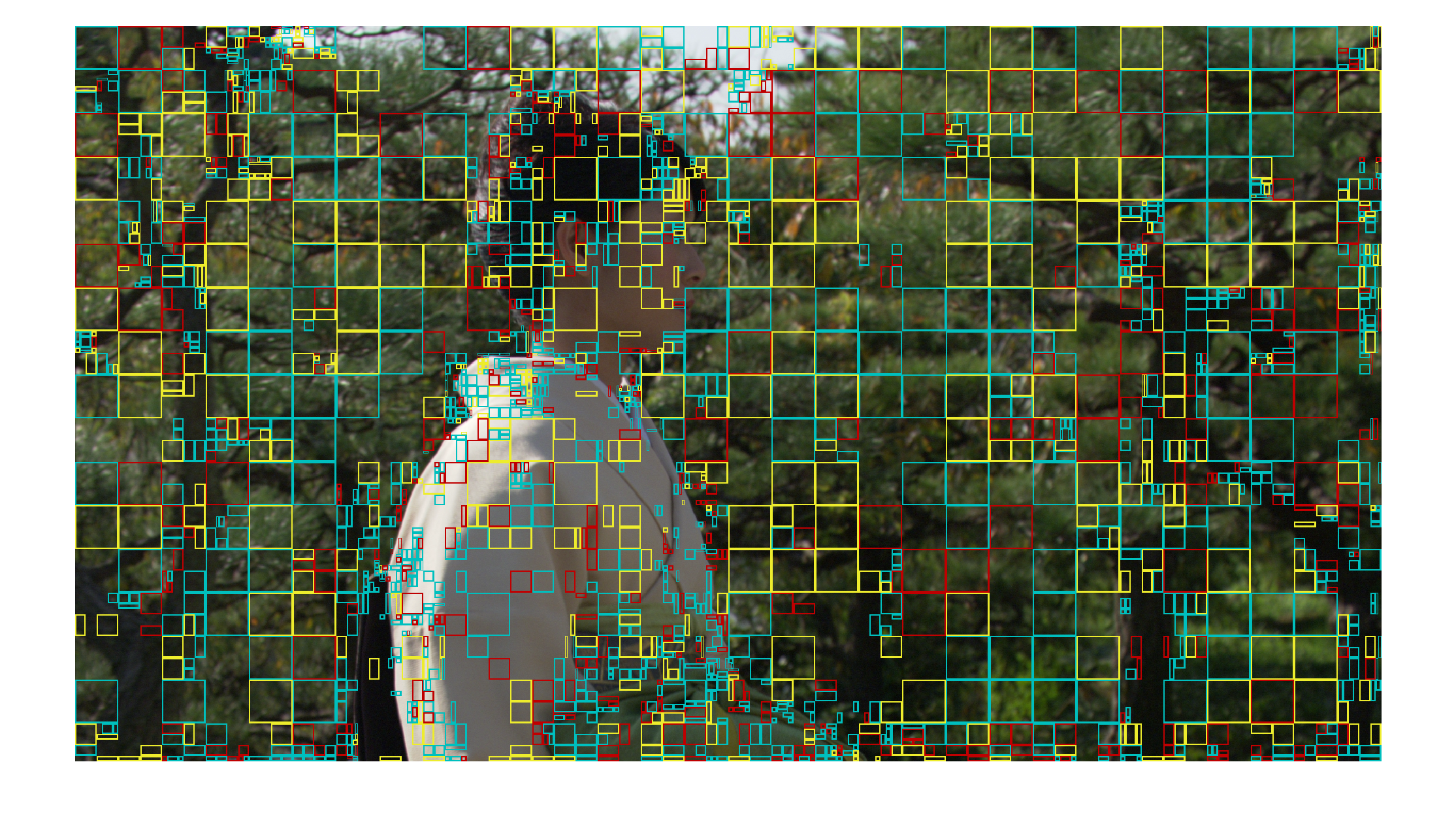}}
\caption{The distribution of blocks coded by fast TreeNet. Yellow, cyan-blue, and red squares represent blocks encoded by networks at level $1$, $2$, and $3$ respectively. The parts not enclosed by squares are encoded by the VVC directional prediction modes.}
\label{fig10}
\end{figure*}

The visual result is shown in Fig. \ref{fig10}. The background parts, such as the tree leaves, are mostly coded by fast TreeNet with large blocks. The foreground parts and background parts with sharp edges, such as the woman's head, the belt, and the trunks, are mostly coded by fast TreeNet with small blocks.

\begin{table}[!t]
\center
\caption{The Usage Rate of Fast TreeNet When Implementing into VTM}
\renewcommand\arraystretch{1.15}
\setlength{\tabcolsep}{6mm}
\scalebox{0.95}{
\begin{tabular}{cc|c}
\hline \hline 
\multicolumn{2}{c|}{\textbf{Sequence}}                                                                                              & \textbf{Usage Rate} \\ \hline
\multicolumn{1}{c|}{\multirow{6}{*}{\textbf{\begin{tabular}[c]{@{}c@{}}Class A\\ (4K)\end{tabular}}}}    & \textbf{Tango2}          & 71.1\%              \\
\multicolumn{1}{c|}{}                                                                                    & \textbf{FoodMarket4}     & 71.7\%              \\
\multicolumn{1}{c|}{}                                                                                    & \textbf{Campfire}        & 61.8\%              \\
\multicolumn{1}{c|}{}                                                                                    & \textbf{CatRobot}        & 48.4\%              \\
\multicolumn{1}{c|}{}                                                                                    & \textbf{DaylightRoad2}   & 51.1\%              \\
\multicolumn{1}{c|}{}                                                                                    & \textbf{ParkRunning3}    & 58.0\%              \\ \hline
\multicolumn{1}{c|}{\multirow{7}{*}{\textbf{\begin{tabular}[c]{@{}c@{}}Class B\\ (1080P)\end{tabular}}}} & \textbf{MarketPlace}     & 54.3\%              \\
\multicolumn{1}{c|}{}                                                                                    & \textbf{RitualDance}     & 53.4\%              \\
\multicolumn{1}{c|}{}                                                                                    & \textbf{Kimono}          & 70.5\%              \\
\multicolumn{1}{c|}{}                                                                                    & \textbf{ParkScene}       & 51.3\%              \\
\multicolumn{1}{c|}{}                                                                                    & \textbf{Cactus}          & 46.1\%              \\
\multicolumn{1}{c|}{}                                                                                    & \textbf{BasketballDrive} & 44.4\%              \\
\multicolumn{1}{c|}{}                                                                                    & \textbf{BQTerrace}       & 39.6\%              \\ \hline
\multicolumn{1}{c|}{\multirow{4}{*}{\textbf{\begin{tabular}[c]{@{}c@{}}ClassC\\ (WVGA)\end{tabular}}}}   & \textbf{BasketballDrill} & 38.5\%              \\
\multicolumn{1}{c|}{}                                                                                    & \textbf{BQMall}          & 44.6\%              \\
\multicolumn{1}{c|}{}                                                                                    & \textbf{PartyScene}      & 49.7\%              \\
\multicolumn{1}{c|}{}                                                                                    & \textbf{RaceHorses}      & 52.3\%              \\ \hline
\multicolumn{1}{c|}{\multirow{4}{*}{\textbf{\begin{tabular}[c]{@{}c@{}}ClassD\\ (WQVGA)\end{tabular}}}}  & \textbf{BasketballPass}  & 38.7\%              \\
\multicolumn{1}{c|}{}                                                                                    & \textbf{BQSquare}        & 38.3\%              \\
\multicolumn{1}{c|}{}                                                                                    & \textbf{BlowingBubbles}  & 51.8\%              \\
\multicolumn{1}{c|}{}                                                                                    & \textbf{RaceHorses}      & 53.6\%              \\ \hline
\multicolumn{1}{c|}{\multirow{3}{*}{\textbf{\begin{tabular}[c]{@{}c@{}}Class E\\ (720P)\end{tabular}}}}  & \textbf{FourPeople}      & 46.1\%              \\
\multicolumn{1}{c|}{}                                                                                    & \textbf{Johnny}          & 50.9\%              \\
\multicolumn{1}{c|}{}                                                                                    & \textbf{KristenAndSara}  & 45.4\%              \\ \hline
\multicolumn{2}{c|}{\textbf{Average}}                                                                                               & 51.3\%   
\\\hline \hline 
\end{tabular}}
\label{table8}
\end{table}

\begin{table}[!t]
\center
\caption{The Model Size of TreeNet}
\renewcommand\arraystretch{1.15}
\setlength{\tabcolsep}{2.3mm}
\scalebox{0.95}{
\begin{tabular}{c|c|c|c|c|c}
\hline \hline 
                    & \textbf{64x64} & \textbf{32x32} & \textbf{16x16} & \textbf{8x8} & \textbf{4x4} \\ \hline
\textbf{Model Size} & 80.5MB         & 48.5MB         & 11.2MB         & 9.5MB        & 2.4 MB      \\\hline \hline 
\end{tabular}}
\label{table7}
\end{table}

\subsection{Model Size and Memory Cost}
The size of each Libtorch model in TreeNet is listed in Table \ref{table7}. The model size for block of 64$\times$64, 32$\times$32, 16$\times$16, 8$\times$8, and 4$\times$4 are 80.5 MB, 48.5 MB, 11.2 MB, 9.5 MB, and 2.4 MB respectively. There are 7 models for each block size in TreeNet, so overall $7\times(80.5+48.5+11.2+9.5+2.4)=1064.7$ MB is needed to store TreeNet with $Depth$=3 in memory. In addition, another $2\times (64+64+8)\times (64+64+8)\div ( 1024)=36.125$KB should be allocated in memory to store the 8-line reference samples.

\begin{table}[!t]
\center
\caption{The Average Training Loss of Each TreeNet Mode at level $2$, $3$, and $4$}
\renewcommand\arraystretch{1.13}
\setlength{\tabcolsep}{1.3mm}
\scalebox{0.95}{
\begin{tabular}{c|c|cccccccc}
\hline\hline  
\multirow{9}{*}{\textbf{\rotatebox{90}{Level 2}}} & \textbf{Iteration} & \multicolumn{4}{c}{\textbf{2}}                                                                       & \multicolumn{4}{c}{\textbf{3}}                                                      \\ \cline{2-10} 
                                   & \textbf{1}         & \multicolumn{4}{c}{124.53}                                                                              & \multicolumn{4}{c}{126.16}                                                             \\
                                   & \textbf{2}         & \multicolumn{4}{c}{113.82}                                                                              & \multicolumn{4}{c}{116.51}                                                             \\
                                   & \textbf{3}         & \multicolumn{4}{c}{109.93}                                                                              & \multicolumn{4}{c}{113.74}                                                             \\
                                   & \textbf{4}         & \multicolumn{4}{c}{108.13}                                                                              & \multicolumn{4}{c}{112.27}                                                             \\
                                   & \textbf{5}         & \multicolumn{4}{c}{107.23}                                                                              & \multicolumn{4}{c}{111.39}                                                             \\
                                   & \textbf{6}         & \multicolumn{4}{c}{106.85}                                                                              & \multicolumn{4}{c}{110.66}                                                             \\
                                   & \textbf{7}         & \multicolumn{4}{c}{106.50}                                                                              & \multicolumn{4}{c}{110.31}                                                             \\
                                   & \textbf{8}         & \multicolumn{4}{c}{105.70}                                                                              & \multicolumn{4}{c}{109.95}                                                             \\ \hline
\multirow{9}{*}{\textbf{\rotatebox{90}{Level 3}}} & \textbf{Iteration} & \multicolumn{2}{c}{\textbf{4}}                  & \multicolumn{2}{c|}{\textbf{5}}                 & \multicolumn{2}{c}{\textbf{6}}                  & \multicolumn{2}{c}{\textbf{7}} \\ \cline{2-10} 
                                   & \textbf{1}         & \multicolumn{2}{c}{101.56}                         & \multicolumn{2}{c|}{90.80}                         & \multicolumn{2}{c}{103.68}                         & \multicolumn{2}{c}{98.84}         \\
                                   & \textbf{2}         & \multicolumn{2}{c}{99.52}                          & \multicolumn{2}{c|}{85.20}                         & \multicolumn{2}{c}{102.26}                         & \multicolumn{2}{c}{92.23}         \\
                                   & \textbf{3}         & \multicolumn{2}{c}{100.11}                         & \multicolumn{2}{c|}{82.73}                         & \multicolumn{2}{c}{101.97}                         & \multicolumn{2}{c}{89.52}         \\
                                   & \textbf{4}         & \multicolumn{2}{c}{101.02}                         & \multicolumn{2}{c|}{81.27}                         & \multicolumn{2}{c}{102.55}                         & \multicolumn{2}{c}{88.07}         \\
                                   & \textbf{5}         & \multicolumn{2}{c}{102.82}                         & \multicolumn{2}{c|}{79.89}                         & \multicolumn{2}{c}{102.63}                         & \multicolumn{2}{c}{87.13}         \\
                                   & \textbf{6}         & \multicolumn{2}{c}{104.15}                         & \multicolumn{2}{c|}{78.93}                         & \multicolumn{2}{c}{102.91}                         & \multicolumn{2}{c}{86.51}         \\
                                   & \textbf{7}         & \multicolumn{2}{c}{105.73}                         & \multicolumn{2}{c|}{77.97}                         & \multicolumn{2}{c}{103.68}                         & \multicolumn{2}{c}{85.39}         \\
                                   & \textbf{8}         & \multicolumn{2}{c}{105.96}                         & \multicolumn{2}{c|}{77.11}                         & \multicolumn{2}{c}{103.94}                         & \multicolumn{2}{c}{85.02}         \\ \hline
\multirow{9}{*}{\textbf{\rotatebox{90}{Level 4}}} & \textbf{Iteration} & \textbf{8} & \multicolumn{1}{c|}{\textbf{9}} & \textbf{10} & \multicolumn{1}{c|}{\textbf{11}} & \textbf{12} & \multicolumn{1}{c|}{\textbf{13}} & \textbf{14}   & \textbf{15}   \\ \cline{2-10} 
                                   & \textbf{1}         & 96.55         & \multicolumn{1}{c|}{97.73}         & 72.47         & \multicolumn{1}{c|}{69.42}         & 93.90         & \multicolumn{1}{c|}{95.52}         & 78.45           & 77.08           \\
                                   & \textbf{2}         & 94.01         & \multicolumn{1}{c|}{91.95}         & 70.47         & \multicolumn{1}{c|}{66.49}         & 88.91         & \multicolumn{1}{c|}{92.71}         & 75.47           & 73.65           \\
                                   & \textbf{3}         & 94.38         & \multicolumn{1}{c|}{88.36}         & 70.01         & \multicolumn{1}{c|}{64.62}         & 86.50         & \multicolumn{1}{c|}{92.45}         & 74.48           & 72.49           \\
                                   & \textbf{4}         & 95.41         & \multicolumn{1}{c|}{85.34}         & 70.63         & \multicolumn{1}{c|}{63.83}         & 84.96         & \multicolumn{1}{c|}{92.23}         & 74.19           & 72.14           \\
                                   & \textbf{5}         & 96.53         & \multicolumn{1}{c|}{83.76}         & 70.33         & \multicolumn{1}{c|}{63.16}         & 83.77         & \multicolumn{1}{c|}{92.66}         & 73.53           & 71.87           \\
                                   & \textbf{6}         & 98.57         & \multicolumn{1}{c|}{81.44}         & 70.54         & \multicolumn{1}{c|}{62.61}         & 82.65         & \multicolumn{1}{c|}{92.76}         & 73.15           & 71.39           \\
                                   & \textbf{7}         & 100.10        & \multicolumn{1}{c|}{80.21}         & 70.40         & \multicolumn{1}{c|}{62.00}         & 82.17         & \multicolumn{1}{c|}{92.95}         & 72.92           & 71.50           \\
                                   & \textbf{8}         & 101.03        & \multicolumn{1}{c|}{78.37}         & 70.88         & \multicolumn{1}{c|}{61.57}         & 81.47         & \multicolumn{1}{c|}{93.04}         & 72.47           & 71.53       \\ \hline\hline   
\end{tabular}
}
\label{table9}
\end{table}

\begin{table}[!t]
\center
\caption{The Rate of Samples Remaining The Same Cluster in Each Iteration and The  Ratio of The Training Data to The Entire Training Dataset after The Last Iteration}
\renewcommand\arraystretch{1.13}
\setlength{\tabcolsep}{1mm}
\scalebox{0.95}{
\begin{tabular}{cc|cccccccc}
\hline\hline 
\multicolumn{1}{c|}{\multirow{8}{*}{\textbf{\rotatebox{90}{Level 2}}}} & \textbf{Iteration} & \multicolumn{4}{c}{\textbf{2}}                                                                       & \multicolumn{4}{c}{\textbf{3}}                                                      \\ \cline{2-10} 
\multicolumn{1}{c|}{}                                   & \textbf{1-2}       & \multicolumn{4}{c}{0.798}                                                                               & \multicolumn{4}{c}{0.793}                                                              \\
\multicolumn{1}{c|}{}                                   & \textbf{2-3}       & \multicolumn{4}{c}{0.914}                                                                               & \multicolumn{4}{c}{0.911}                                                              \\
\multicolumn{1}{c|}{}                                   & \textbf{3-4}       & \multicolumn{4}{c}{0.942}                                                                               & \multicolumn{4}{c}{0.944}                                                              \\
\multicolumn{1}{c|}{}                                   & \textbf{4-5}       & \multicolumn{4}{c}{0.956}                                                                               & \multicolumn{4}{c}{0.957}                                                              \\
\multicolumn{1}{c|}{}                                   & \textbf{5-6}       & \multicolumn{4}{c}{0.965}                                                                               & \multicolumn{4}{c}{0.969}                                                              \\
\multicolumn{1}{c|}{}                                   & \textbf{6-7}       & \multicolumn{4}{c}{0.973}                                                                               & \multicolumn{4}{c}{0.970}                                                              \\
\multicolumn{1}{c|}{}                                   & \textbf{7-8}       & \multicolumn{4}{c}{0.977}                                                                               & \multicolumn{4}{c}{0.974}                                                              \\ \hline
\multicolumn{2}{c|}{\textbf{Ratio}}                                   & \multicolumn{4}{c|}{50.3\%}                                                                             & \multicolumn{4}{c}{49.7\%}                                                             \\ \hline
\multicolumn{1}{c|}{\multirow{8}{*}{\textbf{\rotatebox{90}{Level 3}}}} & \textbf{Iteration} & \multicolumn{2}{c}{\textbf{4}}                  & \multicolumn{2}{c|}{\textbf{5}}                 & \multicolumn{2}{c}{\textbf{6}}                  & \multicolumn{2}{c}{\textbf{7}} \\ \cline{2-10} 
\multicolumn{1}{c|}{}                                   & \textbf{1-2}       & \multicolumn{2}{c}{0.783}                          & \multicolumn{2}{c|}{0.848}                         & \multicolumn{2}{c}{0.786}                          & \multicolumn{2}{c}{0.821}         \\
\multicolumn{1}{c|}{}                                   & \textbf{2-3}       & \multicolumn{2}{c}{0.886}                          & \multicolumn{2}{c|}{0.941}                         & \multicolumn{2}{c}{0.900}                          & \multicolumn{2}{c}{0.925}         \\
\multicolumn{1}{c|}{}                                   & \textbf{3-4}       & \multicolumn{2}{c}{0.916}                          & \multicolumn{2}{c|}{0.966}                         & \multicolumn{2}{c}{0.933}                          & \multicolumn{2}{c}{0.950}         \\
\multicolumn{1}{c|}{}                                   & \textbf{4-5}       & \multicolumn{2}{c}{0.928}                          & \multicolumn{2}{c|}{0.976}                         & \multicolumn{2}{c}{0.946}                          & \multicolumn{2}{c}{0.962}         \\
\multicolumn{1}{c|}{}                                   & \textbf{5-6}       & \multicolumn{2}{c}{0.942}                          & \multicolumn{2}{c|}{0.981}                         & \multicolumn{2}{c}{0.956}                          & \multicolumn{2}{c}{0.969}         \\
\multicolumn{1}{c|}{}                                   & \textbf{6-7}       & \multicolumn{2}{c}{0.945}                          & \multicolumn{2}{c|}{0.984}                         & \multicolumn{2}{c}{0.956}                          & \multicolumn{2}{c}{0.971}         \\
\multicolumn{1}{c|}{}                                   & \textbf{7-8}       & \multicolumn{2}{c}{0.952}                          & \multicolumn{2}{c|}{0.985}                         & \multicolumn{2}{c}{0.962}                          & \multicolumn{2}{c}{0.974}         \\ \hline
\multicolumn{2}{c|}{\textbf{Ratio}}                                   & \multicolumn{2}{c|}{18.6\%}                        & \multicolumn{2}{c|}{31.7\%}                        & \multicolumn{2}{c|}{22.2\%}                        & \multicolumn{2}{c}{27.5\%}        \\ \hline
\multicolumn{1}{c|}{\multirow{8}{*}{\textbf{\rotatebox{90}{Level 4}}}} & \textbf{Iteration} & \textbf{8} & \multicolumn{1}{c|}{\textbf{9}} & \textbf{10} & \multicolumn{1}{c|}{\textbf{11}} & \textbf{12} & \multicolumn{1}{c|}{\textbf{13}} & \textbf{14}   & \textbf{15}   \\ \cline{2-10} 
\multicolumn{1}{c|}{}                                   & \textbf{1-2}       & 0.821         & \multicolumn{1}{c|}{0.834}         & 0.804         & \multicolumn{1}{c|}{0.822}         & 0.822         & \multicolumn{1}{c|}{0.805}         & 0.802           & 0.813           \\
\multicolumn{1}{c|}{}                                   & \textbf{2-3}       & 0.914         & \multicolumn{1}{c|}{0.938}         & 0.906         & \multicolumn{1}{c|}{0.924}         & 0.926         & \multicolumn{1}{c|}{0.907}         & 0.918           & 0.925           \\
\multicolumn{1}{c|}{}                                   & \textbf{3-4}       & 0.934         & \multicolumn{1}{c|}{0.964}         & 0.936         & \multicolumn{1}{c|}{0.951}         & 0.950         & \multicolumn{1}{c|}{0.937}         & 0.949           & 0.951           \\
\multicolumn{1}{c|}{}                                   & \textbf{4-5}       & 0.944         & \multicolumn{1}{c|}{0.973}         & 0.951         & \multicolumn{1}{c|}{0.961}         & 0.963         & \multicolumn{1}{c|}{0.946}         & 0.959           & 0.963           \\
\multicolumn{1}{c|}{}                                   & \textbf{5-6}       & 0.947         & \multicolumn{1}{c|}{0.978}         & 0.957         & \multicolumn{1}{c|}{0.969}         & 0.967         & \multicolumn{1}{c|}{0.958}         & 0.966           & 0.966           \\
\multicolumn{1}{c|}{}                                   & \textbf{6-7}       & 0.950         & \multicolumn{1}{c|}{0.982}         & 0.961         & \multicolumn{1}{c|}{0.972}         & 0.972         & \multicolumn{1}{c|}{0.962}         & 0.968           & 0.970           \\
\multicolumn{1}{c|}{}                                   & \textbf{7-8}       & 0.954         & \multicolumn{1}{c|}{0.983}         & 0.963         & \multicolumn{1}{c|}{0.974}         & 0.976         & \multicolumn{1}{c|}{0.963}         & 0.972           & 0.970           \\ \hline
\multicolumn{2}{c|}{\textbf{Ratio}}                                   & 7.9\%         & \multicolumn{1}{c|}{10.7\%}        & 14.7\%        & \multicolumn{1}{c|}{17.0\%}        & 12.0\%        & \multicolumn{1}{c|}{10.2\%}        & 13.5\%          & 14.0\%         \\\hline\hline 
\end{tabular}
}
\label{table10}
\end{table}

\subsection{Investigations on The Data Clustering-Driven Training}
To assess the data clustering-driven training, we first give some investigations on the prediction losses during the network split and iterative training. The average training losses of each network mode at level $2$, $3$, and $4$ in TreeNet for 8$\times$8 blocks are provided in Tables \ref{table9}. The average training losses of TreeNet modes gradually decrease and converge during the iterative data clustering-driven training. Then we conduct a statistic of the rate of samples remaining in the same cluster after each round of data clustering to see whether the data clustering could finally converge as shown in Table \ref{table10}. In addition, the ratio of the training data to the entire training dataset after the last iteration is also provided. We can observe that less and less training data is clustered to another cluster during the iterative training, which indicates that the data clustering and network training are gradually stabilized.

\section{Conclusions}
In this paper, a novel tree-structured data clustering-driven neural network for intra prediction is proposed which builds the networks and clusters the training data in a tree-structured manner. Specifically, in each network split and training process of TreeNet, each parent network on a leaf node is split into two child networks by adding or subtracting Gaussian random noise. Then a data clustering-driven training is applied to train the two derived child networks. In addition, a fast termination strategy is proposed to accelerate the search of TreeNet. TreeNet could be applied to combine with or replace the directional prediction modes. Although TreeNet has been implemented into HEVC (HM-16.9) and VVC (VTM-4.0) and shows advantages, its network structure is too complex for practical applications. In future, we will focus its simplification like MIP in VVC and study to replace the directional prediction modes in VVC.

\ifCLASSOPTIONcaptionsoff
  \newpage
\fi


bare_
IEEEexample.bib 
V1.12 (2007/01/11)
Copyright (c) 2002-2007 by Michael Shell
See: http://www.michaelshell.org/
for current contact information.

This is an example BibTeX database for the official IEEEtran.bst
BibTeX style file.

Some entries call strings that are defined in the IEEEabrv.bib file.
Therefore, IEEEabrv.bib should be loaded prior to this file. 
Usage: 

\bibliographystyle{./IEEEtran}
\bibliography{./IEEEabrv,./IEEEexample}


Support sites:
http://www.michaelshell.org/tex/ieeetran/
http://www.ctan.org/tex-archive/macros/latex/contrib/IEEEtran/
and/or
http://www.ieee.org/

*************************************************************************
Legal Notice:
This code is offered as-is without any warranty either expressed or
implied; without even the implied warranty of MERCHANTABILITY or
FITNESS FOR A PARTICULAR PURPOSE! 
User assumes all risk.
In no event shall IEEE or any contributor to this code be liable for
any damages or losses, including, but not limited to, incidental,
consequential, or any other damages, resulting from the use or misuse
of any information contained here.

All comments are the opinions of their respective authors and are not
necessarily endorsed by the IEEE.

This work is distributed under the LaTeX Project Public License (LPPL)
( http://www.latex-project.org/ ) version 1.3, and may be freely used,
distributed and modified. A copy of the LPPL, version 1.3, is included
in the base LaTeX documentation of all distributions of LaTeX released
2003/12/01 or later.
Retain all contribution notices and credits.
** Modified files should be clearly indicated as such, including  **
** renaming them and changing author support contact information. **

File list of work: IEEEabrv.bib, IEEEfull.bib, IEEEexample.bib,
                   IEEEtran.bst, IEEEtranS.bst, IEEEtranSA.bst,
                   IEEEtranN.bst, IEEEtranSN.bst, IEEEtran_bst_HOWTO.pdf
*************************************************************************


Note that, because the example references were taken from actual IEEE
publications, these examples do not always contain the full amount
of information that may be desirable (for use with other BibTeX styles).
In particular, full names (not abbreviated with initials) should be
entered whenever possible as some (non-IEEE) bibliography styles use
full names. IEEEtran.bst will automatically abbreviate when it encounters
full names.
 
 
 
 
references for the IEEEtran.bst documentation
IEEEtran homepage
@electronic{IEEEhowto:IEEEtranpage,
  author        = "Michael Shell",
  title         = "{IEEE}tran Homepage",
  url           = "http://www.michaelshell.org/tex/ieeetran/",
  year          = "2007"
}


the distribution site for IEEEtran.bst
@electronic{IEEEexample:shellCTANpage,
  author        = "Michael Shell",
  title         = "{IEEE}tran Webpage on {CTAN}",
  url           = "http://www.ctan.org/tex-archive/macros/latex/contrib/IEEEtran/",
  year          = "2007"
}


the IEEE website
sort key is needed for sorting styles
@electronic{IEEEexample:IEEEwebsite,
  title         = "The {IEEE} Website",
  url           = "http://www.ieee.org/",
  year          = "2007",
  key           = "IEEE"
}


The BibTeX user's guide.
The filename could have been put in the URL instead. But, there might
be other interesting things for the user in the same directory - and
the filename might change before the directory that contains it.
@electronic{IEEEexample:bibtexuser,
  author        = "Oren Patashnik",
  title         = "{{\BibTeX}}ing",
  howpublished  = "{btxdoc.pdf}",
  url           = "http://www.ctan.org/tex-archive/biblio/bibtex/contrib/doc/",
  month         = feb,
  year          = "1988"
}


The BibTeX style designer's guide.
@electronic{IEEEexample:bibtexdesign,
  author        = "Oren Patashnik",
  title         = "Designing {{\BibTeX}} Styles",
  howpublished  = "{btxhak.pdf}",
  url           = "http://www.ctan.org/tex-archive/biblio/bibtex/contrib/doc/",
  month         = feb,
  year          = "1988"
}


A comprehensive BibTeX tutorial.
@electronic{IEEEexample:tamethebeast,
  author        = "Nicolas Markey",
  title         = "Tame the BeaST  ---  The B to X of {{\BibTeX}}",
  url           = "http://tug.ctan.org/tex-archive/info/bibtex/tamethebeast/",
  month         = oct,
  year          = "2005"
}


The BibTeX Tips and FAQ guide.
@electronic{IEEEexample:bibtexFAQ,
  author        = "David Hoadley and Michael Shell",
  title         = "{{\BibTeX}} Tips and {FAQ}",
  howpublished  = "{btxFAQ.pdf}",
  url           = "http://www.ctan.org/tex-archive/biblio/bibtex/contrib/doc/",
  month         = jan,
  year          = "2007"
}


The TeX FAQ
@electronic{IEEEexample:texfaq,
  author        = "Robin Fairbairns",
  title         = "The {{\TeX}} {FAQ}",
  url           = "http://www.tex.ac.uk/cgi-bin/texfaq2html/",
  month         = jan,
  year          = "2007"
}


A BibTeX Guide via Examples.
@electronic{IEEEexample:bibtexguide,
  author        = "Ki-Joo Kim",
  title         = "A {{\BibTeX}} Guide via Examples",
  howpublished  = "{bibtex\_guide.pdf}",
  url           = "http://www.geocities.com/kijoo2000/",
  month         = apr,
  year          = "2004"
}


TeX User Group Bibliography Archive
@electronic{IEEEexample:beebe_archive,
  author        = "Nelson H. F. Beebe",
  title         = "{{\TeX}} User Group Bibliography Archive",
  url           = "http://www.math.utah.edu:8080/pub/tex/bib/index-table.html",
  month         = aug,
  year          = "2006"
}

The natbib.sty package.
@electronic{ctan:natbib,
  author        = "Patrick W. Daly",
  title         = "The natbib.sty package",
  url           = "http://www.ctan.org/tex-archive/macros/latex/contrib/natbib/",
  month         = sep,
  year          = "2006"
}

The url.sty package.
@electronic{IEEEexample:urlsty,
  author        = "Donald Arseneau",
  title         = "The url.sty Package",
  url           = "http://www.ctan.org/tex-archive/macros/latex/contrib/misc/",
  month         = jun,
  year          = "2005",
}


The hyperref.sty package.
@electronic{IEEEexample:hyperrefsty,
  author        = "Sebastian Rahtz and Heiko Oberdiek",
  title         = "The hyperref.sty Package",
  url           = "http://www.ctan.org/tex-archive/macros/latex/contrib/hyperref/",
  month         = nov,
  year          = "2006",
}


The breakurl.sty package.
@electronic{IEEEexample:breakurl,
  author        = "Vilar Camara Neto",
  title         = "The breakurl.sty Package",
  url           = "http://www.ctan.org/tex-archive/macros/latex/contrib/breakurl/",
  month         = aug,
  year          = "2006",
}


The Babel package.
@electronic{IEEEexample:babel,
  author        = "Johannes Braams",
  title         = "The Babel Package",
  url           = "http://www.ctan.org/tex-archive/macros/latex/required/babel/",
  month         = nov,
  year          = "2005",
}


The multibib package.
@electronic{IEEEexample:multibibsty,
  author        = "Thorsten Hansen",
  title         = "The multibib.sty package",
  url           = "http://www.ctan.org/tex-archive/macros/latex/contrib/multibib/",
  month         = jan,
  year          = "2004"
}


The biblatex package.
@electronic{IEEEexample:biblatex,
  author        = "Philipp Lehman",
  title         = "The biblatex package",
  url           = "http://www.ctan.org/tex-archive/macros/latex/exptl/biblatex/",
  month         = jan,
  year          = "2007"
}



The three most common and typical types of references used in
IEEE publications:

an article in a journal
Note the use of the IEEE_J_EDL string, defined in the IEEEabrv.bib file,
for the journal name. IEEEtran.bst defines the BibTeX standard three
letter month codes per IEEE style.
From the June 2002 issue of "IEEE Transactions on Electron Devices",
page 996, reference #16.
@article{IEEEexample:article_typical,
  author        = "S. Zhang and C. Zhu and J. K. O. Sin and P. K. T. Mok",
  title         = "A Novel Ultrathin Elevated Channel Low-temperature 
                   Poly-{Si} {TFT}",
  journal       = IEEE_J_EDL,
  volume        = "20",
  month         = nov,
  year          = "1999",
  pages         = "569-571"
}


journal article using et al.
The (five) authors are actually: F. Delorme, S. Slempkes, G. Alibert, 
B. Rose, J. Brandon
The month (July) was not given here.
From the September 1998 issue of "IEEE Journal on Selected Areas in
Communications", page 1257, reference #28.
@article{IEEEexample:articleetal,
  author        = "F. Delorme and others",
  title         = "Butt-jointed {DBR} Laser With 15 {nm} Tunability Grown
                   in Three {MOVPE} Steps",
  journal       = "Electron. Lett.",
  volume        = "31",
  number        = "15",
  year          = "1995",
  pages         = "1244-1245"
}


a paper in a conference proceedings
"conference" can be used as an alias for "inproceedings"
From the June 2002 issue of "Journal of Microelectromechanical Systems",
page 205, reference #16.
@inproceedings{IEEEexample:conf_typical,
  author        = "R. K. Gupta and S. D. Senturia",
  title         = "Pull-in Time Dynamics as a Measure of Absolute Pressure",
  booktitle     = "Proc. {IEEE} International Workshop on
                   Microelectromechanical Systems ({MEMS}'97)",
  address       = "Nagoya, Japan",
  month         = jan,
  year          = "1997",
  pages         = "290-294"
}


a book
From the May 2002 issue of "IEEE Transactions on Magnetics",
page 1466, reference #4.
@book{IEEEexample:book_typical,
  author        = "B. D. Cullity",
  title         = "Introduction to Magnetic Materials",
  publisher     = "Addison-Wesley",
  address       = "Reading, MA",
  year          = "1972"
}




Other examples

journal article with large page numbers, IEEE will divide numbers 5 digits
or longer into groups of three with small spaces between them. Page ranges
can be separated via either "-" or "--", IEEEtran.bst will automatically
convert the separator dash(es) to "--".
Authors and/or IEEE do not always provide/use the journal number, but it
was used in this case. IEEEtran.bst can be configured to ignore journal
numbers if desired.
From the August 2000 issue of "IEEE Photonics Technology Letters",
page 1039, reference #11.
@article{IEEEexample:articlelargepages,
  author        = "A. Castaldini and A. Cavallini and B. Fraboni
                   and P. Fernandez and J. Piqueras",
  title         = "Midgap Traps Related to Compensation Processes in
                   {CdTe} Alloys",
  journal       = "Phys. Rev. B.",
  volume        = "56",
  number        = "23",
  year          = "1997",
  pages         = "14897-14900"
}


journal article with dual months 
use the BibTeX "#" concatenation operator
From the March 2002 issue of "IEEE Transactions on Mechatronics",
page 21, reference #8.
@article{IEEEexample:articledualmonths,
  author        = "Y. Okada and K. Dejima and T. Ohishi",
  title         = "Analysis and Comparison of {PM} Synchronous Motor and
                   Induction Motor Type Magnetic Bearings",
  journal       = IEEE_J_IA,
  volume        = "31",
  month         = sep # "/" # oct,
  year          = "1995",
  pages         = "1047-1053"
}


journal article to be published as a misc entry type
date information like month and year is optional 
However, the article form like that below may be a better approach.
From the May 2002 issue of "IEEE Journal of Selected Areas in 
Communication", page 725, reference #3.
@misc{IEEEexample:TBPmisc,
  author        = "M. Coates and A. Hero and R. Nowak and B. Yu",
  title         = "Internet Tomography",
  howpublished  = IEEE_M_SP,
  month         = may,
  year          = "2002",
  note          = "to be published"
}


journal article to be published as an article entry type
year is required, so if absent, use the year field to hold
the "submitted for publication" in order to avoid a warning for
the missing year field.
From the June 2002 issue of "IEEE Transactions on Information Theory",
page 1461, reference #21.
@article{IEEEexample:TBParticle,
  author        = "N. Kahale and R. Urbanke",
  title         = "On the Minimum Distance of Parallel and Serially
                   Concatenated Codes",
  journal       = IEEE_J_IT,
  year          = "submitted for publication"
}





book with editor and no author
From the June 2002 issue of "IEEE Transactions on Information Theory",
page 1725, reference #1.
@book{IEEEexample:bookwitheditor,
  editor        = "J. C. Candy and G. C. Temes",
  title         = "Oversampling Delta-Sigma Data Converters Theory,
                   Design and Simulation",
  publisher     = "{IEEE} Press.",
  address       = "New York",
  year          = "1992"
}


book with edition, author and editor
Note that the standard BibTeX styles do not support book entries with both
author and editor fields, but IEEEtran.bst does.
The standard BibTeX way of specifying the edition is to use capitalized
ordinal words such as "First", "Second", etc. Most .bst files can convert
up to about "Fifth" into the format needed. IEEEtran.bst can convert up
to "Tenth" to the "Arabic ordinal" form IEEE uses (e.g., "10th"). For
editions over the tenth, it is best to use the "Arabic ordinal" form for
IEEE related work (e.g., "101st")
Note how "Jr." has to be entered.
From the May 2002 issue of "Journal of Lightwave Technology", page 856,
reference #17.
@book{IEEEexample:book,
  author        = "S. M. Metev and V. P. Veiko",
  editor        = "Osgood, Jr., R. M.",
  title         = "Laser Assisted Microtechnology",
  edition       = "Second",
  publisher     = "Springer-Verlag",
  address       = "Berlin, Germany",
  year          = "1998"
}


book with series and volume
From the January 2000 issue of "IEEE Transactions on Communications",
page 11, reference #31.
@book{IEEEexample:bookwithseriesvolume,
  editor        = "J. Breckling",
  title         = "The Analysis of Directional Time Series: Applications to
                   Wind Speed and Direction",
  series        = "Lecture Notes in Statistics",
  publisher     = "Springer",
  address       = "Berlin, Germany",
  year          = "1989",
  volume        = "61"
}


inbook with chapter number. The pages field could also have been given.
The chapter number could be changed to something else such as a section
number via the type field: type = "sec.".
From the May 2002 issue of "IEEE Transactions on Circuits and Systems---I: 
Fundamental Applications and Theory", page 638, reference #22.
@inbook{IEEEexample:inbook,
  author        = "H. E. Rose",
  title         = "A Course in Number Theory",
  publisher     = "Oxford Univ. Press",
  address       = "New York, NY",
  year          = "1988",
  chapter       = "3"
}


inbook with pages and note. The language field is not set to Russian
because the title is presented here in its translated, English, form.
From the May 2002 issue of "IEEE Transactions on Magnetics", page 1533,
reference #5.
@inbook{IEEEexample:inbookpagesnote,
  author        = "B. K. Bul",
  title         = "Theory Principles and Design of Magnetic Circuits",
  publisher     = "Energia Press",
  address       = "Moscow",
  year          = "1964",
  pages         = "464",
  note          = "(in Russian)"
}





incollection with author and editor
From the May 2002 issue of "Journal of Lightwave Technology",
page 807, reference #7.
@incollection{IEEEexample:incollection,
  author        = "W. V. Sorin",
  editor        = "D. Derickson",
  title         = "Optical Reflectometry for Component Characterization",
  booktitle     = "Fiber Optic Test and Measurement",
  publisher     = "Prentice-Hall",
  address       = "Englewood Cliffs, NJ",
  year          = "1998"
}


incollection with series
From the April 2000 issue of "IEEE Transactions on Communication",
page 609, reference #3.
@incollection{IEEEexample:incollectionwithseries,
  author        = "J. B. Anderson and K. Tepe",
  title         = "Properties of the Tailbiting {BCJR} Decoder",
  booktitle     = "Codes, Systems and Graphical Models",
  series        = "{IMA} Volumes in Mathematics and Its Applications",
  publisher     = "Springer-Verlag",
  address       = "New York",
  year          = "2000"
  
}


incollection with author, editor, chapter and pages
From the January 2000 issue of "IEEE Transactions on Communications",
page 16, reference #9.
@incollection{IEEEexample:incollection_chpp,
  author        = "P. Hedelin and P. Knagenhjelm and M. Skoglund",
  editor        = "W. B. Kleijn and K. K. Paliwal",
  title         = "Theory for Transmission of Vector Quantization Data",
  booktitle     = "Speech Coding and Synthesis",
  publisher     = "Elsevier Science",
  address       = "Amsterdam, The Netherlands",
  year          = "1995",
  chapter       = "10",
  pages         = "347-396"
}


incollection with a large number of authors, some authors/journals will
use et al. for so many names. IEEEtran.bst can be configured to do this
if desired, or "R. M. A. Dawson and others" can be used instead.
Note that IEEE may not include the publisher for incollection entries -
IEEEtran.bst will not issue a warning if the publisher is missing for
incollections - but other .bst files often will.
From the June 2002 issue of "IEEE Transactions on Electron Devices",
page 996, reference #12.
@incollection{IEEEexample:incollectionmanyauthors,
  author        = "R. M. A. Dawson and Z. Shen and D. A. Furst and
                   S. Connor and J. Hsu and M. G. Kane and R. G. Stewart and
                   A. Ipri and C. N. King and P. J. Green and R. T. Flegal
                   and S. Pearson and W. A. Barrow and E. Dickey and K. Ping
                   and C. W. Tang and S. Van. Slyke and
                   F. Chen and J. Shi and J. C. Sturm and M. H. Lu",
  title         = "Design of an Improved Pixel for a Polysilicon 
                   Active-Matrix Organic {LED} Display",
  booktitle     = "{SID} Tech. Dig.",
  volume        = "29",
  year          = "1998",
  pages         = "11-14"
}





A Motorola data book as a manual
It is somewhat unusual to include the data book part number.
in the title. It might be more correct to put this information
in the howpublished field instead.
From the December 2000 issue of "IEEE Transactions on Communications",
page 1986, reference #10.
@manual{IEEEexample:motmanual,
  title         = "{FLEXChip} Signal Processor ({MC68175/D})",
  organization  = "Motorola",
  year          = "1996"
}


same reference, but using IEEEtran's howpublished extension
@manual{IEEEexample:motmanualhowpub,
  title         = "{FLEXChip} Signal Processor",
  howpublished  = "{MC68175/D}",
  organization  = "Motorola",
  year          = "1996"
}




conference paper with an address and days. Some journals capitalize the
letters in "Globecom", this one did not.
From the May 2002 issue of "IEEE Transactions on Communications",
page 697, reference #12.
@inproceedings{IEEEexample:confwithadddays,
  author        = "M. S. Yee and L. Hanzo",
  title         = "Radial Basis Function Decision Feedback Equaliser
                   Assisted Burst-by-burst Adaptive Modulation",
  booktitle     = "Proc. {IEEE} Globecom '99",
  address       = "Rio de Janeiro, Brazil",
  month         = dec # " 5--9,",
  year          = "1999",
  pages         = "2183-2187"
}


conference paper with volume number. A conference proceedings with a vol
number is a little uncommon, note that the vol number is placed
before the address in the formatted bibliography entry
From the April 2002 issue of "IEEE/ACM Transactions on Networking",
page 181, reference #26.
@inproceedings{IEEEexample:confwithvolume,
  author        = "M. Yajnik and S. B. Moon and J. Kurose and D. Towsley",
  title         = "Measurement and Modeling of the Temporal Dependence in
                   Packet Loss",
  booktitle     = "Proc. {IEEE} {INFOCOM}'99",
  volume        = "1",
  address       = "New York, NY",
  month         = mar,
  year          = "1999",
  pages         = "345-352"
}


conference paper with a paper number, the type field can be used to
override the word "paper", e.g., type = "postdeadline paper". A type
can be given even without a paper  field.
Also, IEEEtran.bst can be configured to ignore paper numbers and types.
From the May 2002 issue of "Journal of Lightwave Technology",
page 807, reference #4.
@inproceedings{IEEEexample:confwithpaper,
  author        = "M. Wegmuller and J. P. von der Weid and P. Oberson
                   and N. Gisin",
  title         = "High Resolution Fiber Distributed Measurements With
                   Coherent {OFDR}",
  booktitle     = "Proc. {ECOC}'00",
  year          = "2000",
  paper         = "11.3.4",
  pages         = "109"
}


conference paper with a postdeadline type paper, the paper field is
optional.
From the August 2000 issue of "IEEE Photonics Technology Letters",
page 1087, reference #12.
@inproceedings{IEEEexample:confwithpapertype,
  author        = "B. Mikkelsen and G. Raybon and R.-J. Essiambre and
                   K. Dreyer and Y. Su. and L. E. Nelson and J. E. Johnson
                   and G. Shtengel and A. Bond and D. G. Moodie and
                   A. D. Ellis",
  title         = "160 {Gbit/s} Single-channel Transmission Over 300 km 
                   Nonzero-dispersion Fiber With Semiconductor Based
                   Transmitter and Demultiplexer",
  booktitle     = "Proc. {ECOC}'99",
  year          = "1999",
  paper         = "2-3",
  type          = "postdeadline paper",
  pages         = "28-29"
}


presented at a conference
intype overrides the default "in" and causes the booktitle not to be
emphasized (rendered in italics).
From the February 2002 issue of "IEEE/ACM Transactions on Networking",
page 163, reference #6.
@inproceedings{IEEEexample:presentedatconf,
  author        = "S. G. Finn and M. M{\'e}dard and R. A. Barry",
  title         = "A Novel Approach to Automatic Protection Switching
                   Using Trees",
  intype        = "presented at the",
  booktitle     = "Proc. Int. Conf. Commun.",
  year          = "1997"
}





master's thesis, often the University name will be abbreviated and the
state or country will be included in the address. The type field can
used to override the default type "Master's thesis"
From the June 2002 issue of "IEEE Transactions on Microelectromechanical
Systems", page 186, reference #11.
@mastersthesis{IEEEexample:masters,
  author        = "Nin C. Loh",
  title         = "High-Resolution Micromachined Interferometric
                   Accelerometer",
  school        = "Massachusetts Institute of Technology",
  address       = "Cambridge",
  year          = "1992"
}


master's thesis with a type field
From the August 2001 issue of "IEEE/ACM Transactions on Networking",
page 391, reference #12.
@mastersthesis{IEEEexample:masterstype,
  author        = "A. Karnik",
  title         = "Performance of {TCP} Congestion Control with Rate
                   Feedback: {TCP/ABR} and Rate Adaptive {TCP/IP}",
  school        = "Indian Institute of Science",
  type          = "M. Eng. thesis",
  address       = "Bangalore, India",
  month         = jan,
  year          = "1999"
}





Ph.D. dissertation with a URL field, the university is abbreviated
From the October 2001 issue of "IEEE/ACM Transactions on Networking",
page 590, reference #11.
@phdthesis{IEEEexample:phdurl,
  author        = "Q. Li",
  title         = "Delay Characterization and Performance Control of
                   Wide-area Networks",
  school        = "Univ. of Delaware",
  address       = "Newark",
  month         = may,
  year          = "2000",
  url           = "http://www.ece.udel.edu/~qli"
}





technical report
From the August 2001 issue of "IEEE/ACM Transactions on Networking",
page 490, reference #15.
@techreport{IEEEexample:techrep,
  author        = "R. Jain and K. K. Ramakrishnan and D. M. Chiu",
  title         = "Congestion Avoidance in Computer Networks with a 
                   Connectionless Network Layer",
  institution   = "Digital Equipment Corporation",
  address       = "MA",
  number        = "DEC-TR-506",
  month         = aug,
  year          = "1987"
}


technical report with type
for those times when "Tech. Rep." needs to be modified
From the February 2001 issue of "IEEE/ACM Transactions on Networking",
page 46, reference #8.
@techreport{IEEEexample:techreptype,
  author        = "J. Padhye and V. Firoiu and D. Towsley",
  title         = "A Stochastic Model of {TCP} {R}eno Congestion Avoidance
                   and Control",
  institution   = "Univ. of Massachusetts",
  address       = "Amherst, MA",
  type          = "CMPSCI Tech. Rep.",
  number        = "99-02",
  year          = "1999"
}


technical report with type
for those times when "Tech. Rep." needs to be modified
This reference did not have an address.
From the January 2000 issue of "IEEE Transactions on Communications",
page 117, reference #6.
@techreport{IEEEexample:techreptypeii,
  author        = "D. Middleton and A. D. Spaulding",
  title         = "A Tutorial Review of Elements of Weak Signal Detection
                   in Non-{G}aussian {EMI} Environments",
  institution   = "National Telecommunications and Information
                   Administration ({NTIA}), U.S. Dept. of Commerce",
  type          = "NTIA Report",
  number        = "86-194",
  month         = may,
  year          = "1986"
}





an unpublished work
for unpublished types, the note field is required. IEEE usually
just uses the word "unpublished" for the note.
From the August 2001 issue of "IEEE/ACM Transactions on Networking",
page 391, reference #16.
@unpublished{IEEEexample:unpublished,
  author        = "T. J. Ott and N. Aggarwal",
  title         = "{TCP} over {ATM}: {ABR} or {UBR}",
  note          = "Unpublished"
}





electronic with a howpublished information field 
From the August 2001 issue of "IEEE/ACM Transactions on Networking",
page 391, reference #7.
@electronic{IEEEexample:electronhowinfo,
  author        = "V. Jacobson",
  title         = "Modified {TCP} Congestion Avoidance Algorithm",
  howpublished  = "end2end-interest mailing list",
  url           = "ftp://ftp.isi.edu/end2end/end2end-interest-1990.mail",
  month         = apr,
  year          = "1990"
}


electronic with a howpublished information field 
From the August 2001 issue of "IEEE/ACM Transactions on Networking",
page 418, reference #31.
@electronic{IEEEexample:electronhowinfo2,
  author        = "V. Valloppillil and K. W. Ross",
  title         = "Cache Array Routing Protocol v1.1",
  howpublished  = "Internet draft",
  url           = "http://ds1.internic.net/internet-drafts/draft-vinod-carp-v1-03.txt",
  year          = "1998"
}


electronic with an organization and address
From the February 2002 issue of "IEEE/ACM Transactions on Networking",
page 114, reference #15.
@electronic{IEEEexample:electronorgadd,
  author        = "D. H. Lorenz and A. Orda",
  title         = "Optimal Partition of {QoS} Requirements on Unicast
                   Paths and Multicast Trees",
  organization  = "Dept. Elect. Eng., Technion",
  address       = "Haifa, Israel",
  url           = "ftp://ftp.technion.ac.il/pub/supported/ee/Network/lor.mopq98.ps",
  month         = jul,
  year          = "1998"
}





U.S. patent
Use the type field to override the patent type. e.g.,
type = "Patent Application"
The address is that of the assignee. Note that IEEE does not
display the assignee, the address, and only displays one date.
(if year is not present, the filed dates are used.) However, this
information should be entered as other BibTeX styles may use it.
alternatively, nationality could be entered as "U.S." 
From the April 2000 issue of "IEEE Transactions on Communications",
page 542, reference #6.
@patent{IEEEexample:uspat,
  author        = "Ronald E. Sorace and Victor S. Reinhardt and
                   Steven A. Vaughn",
  assignee      = "Hughes Aircraft Company",
  address       = "Los Angeles, CA",
  title         = "High-Speed Digital-to-{RF} Converter",
  nationality   = "United States",
  number        = "5668842",
  dayfiled      = "28",
  monthfiled    = feb,
  yearfiled     = "1995",
  day           = "16",
  month         = sep,
  year          = "1997"
}


Japanese Patent
From the April 2000 issue of "IEEE Transactions on Communications",
page 556, reference #6.
@patent{IEEEexample:jppat,
  author        = "U. Hideki",
  title         = "Quadrature Modulation Circuit",
  nationality   = "Japanese",
  number        = "152932/92",
  day           = "20",
  month         = may,
  year          = "1992"
}


French Patent request, the language field must be entered in lower case
as this is the option name Babel uses. The nationality field needs to be
capitalized. Because this is a patent request, the date filed fields are
used while the date fields are left empty/missing. In other countries,
the words "Patent Application", etc. are used instead.
From the April 2000 issue of "IEEE Transactions on Communications",
page 556, reference #9.
@patent{IEEEexample:frenchpatreq,
  author        = "F. Kowalik and M. Isard",
  title         = "Estimateur d'un D{\'e}faut de Fonctionnement 
                   d'un Modulateur en Quadrature et {\'E}tage de Modulation
                   l'Utilisant",
  language      = "french",
  nationality   = "French",
  type          = "Patent Request",
  number        = "9500261",
  dayfiled      = "11",
  monthfiled    = jan,
  yearfiled     = "1995"
}





a periodical
From the April 2001 issue of "IEEE/ACM Transactions on Networking",
page 160, reference #1.
sort key is needed for sorting styles
@periodical{IEEEexample:periodical,
  title         = IEEE_M_PCOM # ", Special Issue on Wireless {ATM}",
  volume        = "3",
  month         = aug,
  year          = "1996",
  key           = "IEEE"
}





standard, IEEE does not use the address for standards, but it is good
to provide one for BibTeX styles that use it.
From the August 2001 issue of "IEEE/ACM Transactions on Networking",
page 451, reference #2.
@standard{IEEEexample:standard,
  title         = "Wireless {LAN} Medium Access Control {(MAC)} and 
                   Physical Layer {(PHY)} Specification",
  organization  = "IEEE",
  address       = "Piscataway, NJ",
  number        = "802.11",
  year          = "1997"
}


standard with type and revision, the type overrides the word standard
(or std.). Here a standard number is not available and a revision number
is used.
From the August 2000 issue of "IEEE Photonics Technology Letters",
page 1048, reference #13.
@standard{IEEEexample:standardproposed,
  title         = "Fiber Channel Physical Interface ({FC-PI})",
  institution   = "NCITS",
  address       = "Washington, DC",
  type          = "Working Draft Proposed Standard",
  revision      = "5.2",
  year          = "1999"
}


standard draft as a misc with author
From the May 2002 issue of "IEEE Journal of Selected Areas in
Communication", page 725, reference #16.
@misc{IEEEexample:draftasmisc,
  author        = "I. Widjaja and A. Elwalid",
  title         = "{MATE}: {MPLS} Adaptive Traffic Engineering",
  howpublished  = "IETF Draft",
  year          = "1999"
}





misc for a techreport like reference
techreport is not perfectly suitable because this entry lacks
an institution field
From the August 2001 issue of "IEEE/ACM Transactions on Networking",
page 490, reference #22.
@misc{IEEEexample:miscforum,
  author        = "L. Roberts",
  title         = "Enhanced Proportional Rate Control Algorithm {PRCA}",
  howpublished  = "{ATM} Forum Contribution 94-0735R1",
  month         = aug,
  year          = "1994"
}


misc for a white paper
From the August 2001 issue of "IEEE/ACM Transactions on Networking",
page 478, reference #4 - Note that the reference there (improperly?)
used the author field for "Cisco".
@misc{IEEEexample:whitepaper,
  title         = "Advanced {QoS} Services for the Intelligent Internet",
  howpublished  = "White Paper",
  organization  = "Cisco",
  month         = may,
  year          = "1997"
}


misc for a data sheet
From the November 2000 issue of "IEEE Photonics Technology Letters",
page 1551, reference #6.
@misc{IEEEexample:datasheet,
  title         = "{PDCA12-70} Data Sheet",
  organization  = "Opto Speed SA",
  address       = "Mezzovico, Switzerland"
}





Other unusual references

a private communication as a misc entry type
sometimes the designation "personal communication" is used instead
From the June 2002 issue of "IEEE Transactions on Information Theory",
page 1725, reference #16.
@misc{IEEEexample:private,
  author        = "S. Konyagin",
  howpublished  = "private communication",
  year          = "1998"
}


an internet request for comments (RFC) as a misc entry type
It would also be nice to add a URL to these types of things.
RFCs can also be handled as electronic references.
From the April 2002 issue of "IEEE/ACM Transactions on Networking",
page 181, reference #14.
@misc{IEEEexample:miscrfc,
  author        = "K. K. Ramakrishnan and S. Floyd",
  title         = "A Proposal to Add Explicit Congestion
                   Notification ({ECN}) to {IP}",
  howpublished  = "RFC 2481",
  month         = jan,
  year          = "1999"
}


a software package as a manual
From the June 2002 issue of "IEEE/ASME Journal of Microelectromechanical
Systems", page 205, reference #20.
Sometimes they put the version/release information in the title.
@manual{IEEEexample:softmanual,
  title         = "SaberDesigner Reference Manual",
  organization  = "Analogy, Inc.",
  address       = "Beaverton, OR",
  year          = "1998",
  note          = "Release 4.3"
}


a software package as an electronic reference
From the February 2003 issue of  "IEEE/ACM Transactions on Networking",
page 46, reference #24. If there is no author or organization, a sorting
key is required for sorting styles. It might be a good idea to include
month and year fields as well.
@electronic{IEEEexample:softonline,
  title         = "Ucb/lbnl/vint Network Simulator---ns (Version 2)",
  url           = "http://www-mash.cs.berkeley.edu/ns/",
  key           = "ns"
}


misc for a German regulation
In German, the first letters of nouns are capitalized, so we do so here.
From the June 2002 issue of "IEEE Journal in Selected Areas in
Communication", page 892, reference #9.
@misc{IEEEexample:miscgermanreg,
  title         = "{M}essung von {S}t{\"o}rfeldern an {A}nlagen 
                   und {L}eitungen der {T}elekommunikation im
                   {F}requenzbereich 9 {kHz} bis 3 {GHz}",
  language      = "german",
  howpublished  = "{M}e{\ss}vorschrift {R}eg {TP} {MV} 05",
  organization  = "Regulierungsbeh{\"o}rde f{\"u}r {T}elekommunikation und
                   {P}ost ({R}eg {TP})"
}



Ways to handle things like CCSDS's Blue Books
journal article with a URL. However, this is not a very good approach
because the Blue Books are not really journals and the author field has
to be abused.
From the June 2002 issue of "IEEE Transactions on Information Theory",
page 1461, reference #7.
@article{IEEEexample:bluebookarticle,
  author        = "{Consulative Committee for Space Data Systems (CCSDS)}",
  title         = "Telemetry Channel Coding",
  journal       = "Blue Book",
  number        = "4",
  year          = "1999",
  url           = "http://www.ccsds.org/documents/pdf/CCSDS-101.0-B-4.pdf"
}


CCSDS's Blue Book handled as a book
However, it is not a good idea to have to use the author field for
an organization (done here because the book entry type requires an
author field).
@book{IEEEexample:bluebookbook,
  author        = "{Consulative Committee for Space Data Systems (CCSDS)}",
  title         = "Telemetry Channel Coding",
  series        = "Blue Book",
  number        = "4",
  publisher     = "{CCSDS}",
  address       = "Newport Beach, {CA}",
  year          = "1999",
  url           = "http://www.ccsds.org/documents/pdf/CCSDS-101.0-B-4.pdf"
}


CCSDS's Blue Book handled as a manual
This is a much better approach, but uses the howpublished field.
@manual{IEEEexample:bluebookmanual,
  title         = "Telemetry Channel Coding",
  howpublished  = "ser. Blue Book, No. 4",
  organization  = "Consulative Committee for Space Data Systems (CCSDS)",
  address       = "Newport Beach, CA",
  year          = "1999",
  url           = "http://www.ccsds.org/documents/pdf/CCSDS-101.0-B-4.pdf"
}



CCSDS's Blue Book handled as a standard
Probably the best approach for this particular case because the work
is standard related. Note that IEEE does not display the address for
standards.
@standard{IEEEexample:bluebookstandard,
  title         = "Telemetry Channel Coding",
  howpublished  = "ser. Blue Book, No. 4",
  organization  = "Consulative Committee for Space Data Systems (CCSDS)",
  address       = "Newport Beach, CA",
  type          = "Recommendation for Space Data System Standard",
  number        = "101.0-B-4",
  month         = May,
  year          = "1999",
  url           = "http://www.ccsds.org/documents/pdf/CCSDS-101.0-B-4.pdf"
}








An example of a IEEEtran control entry which can change some IEEEtran.bst
settings. An entry like this must be cited via \bstctlcite{} command
before the first real \cite{}. The same entry key cannot be called twice
(just like multiple \cite{} of the same entry key place only one entry
in the bibliography.)
The available control fields are:

CTLuse_article_number
"no" turns off the display of the number for articles.
"yes" enables

CTLuse_paper
"no" turns off the display of the paper and type fields in inproceedings.
"yes" enables

CTLuse_forced_etal 
"no" turns off the forced use of "et al."
"yes" enables

CTLmax_names_forced_etal
The maximum number of names that can be present beyond which an "et al."
usage is forced. Be sure that CTLnames_show_etal (below)
is not greater than this value!

CTLnames_show_etal
The number of names that will be shown with a forced "et al.".
Must be less than or equal to CTLmax_names_forced_etal

CTLuse_alt_spacing 
"no" turns off the alternate interword spacing for entries with URLs.
"yes" enables

CTLalt_stretch_factor
If alternate interword spacing for entries with URLs is enabled, this is
the interword spacing stretch factor that will be used. For example, the
default "4" here means that the interword spacing in entries with URLs can
stretch to four times normal. Does not have to be an integer.

CTLdash_repeated_names
"no" turns off the "dashification" of repeated (i.e., identical to those
of the previous entry) names. IEEE normally does this.
"yes" enables

CTLname_format_string
The name format control string as explained in the BibTeX style hacking
guide.
IEEE style "{f.~}{vv~}{ll}{, jj}" is the default,

CTLname_latex_cmd
A LaTeX command that each name will be fed to (e.g., "\textsc").
Leave empty if no special font is desired for the names.
The default is empty.

CTLname_url_prefix
The prefix text used before URLs.
The default is "[Online]. Available:" A space will be inserted after this
text. If this space is not wanted, just use \relax at the end of the
prefix text.


Those fields that are not to be changed can be left out.
@IEEEtranBSTCTL{IEEEexample:BSTcontrol,
  CTLuse_article_number     = "yes",
  CTLuse_paper              = "yes",
  CTLuse_forced_etal        = "no",
  CTLmax_names_forced_etal  = "10",
  CTLnames_show_etal        = "1",
  CTLuse_alt_spacing        = "yes",
  CTLalt_stretch_factor     = "4",
  CTLdash_repeated_names    = "yes",
  CTLname_format_string     = "{f.~}{vv~}{ll}{, jj}",
  CTLname_latex_cmd         = "",
  CTLname_url_prefix        = "[Online]. Available:"
}




\begin{thebibliography}{1}
\bibitem{b1} \emph{Versatile Video Coding}, document ITU-T Rec. H.266 and ISO/IEC
23090-3, ITU-T and ISO/IEC, 2020.
\bibitem{b2} \emph{High Efficiency Video Coding}, document ITU-T Rec. H.265 and ISO/IEC 23008-2, vers. 1, ITU-T and ISO/IEC, 2013.
\bibitem{b3} G. J. Sullivan, J. R. Ohm, W. J. Han, and T. Wiegand, “Overview of the high efficiency video coding (HEVC) standard,” \emph{IEEE Transactions on Circuits and Systems for Video Technology}, vol. 22, no. 12, pp. 1649–1668, Dec 2012.
\bibitem{b4} B. Bross \emph{et al.}, “Overview of the Versatile Video Coding (VVC) Standard and its Applications,” \emph{IEEE Transactions on Circuits and Systems for Video Technology}, vol. 31, no. 10, pp. 3736-3764, Oct 2021.
\bibitem{VVCintra} J. Pfaff \emph{et al.}, “Intra Prediction and Mode Coding in VVC,” \emph{IEEE Transactions on Circuits and Systems for Video Technology}, vol. 31, no. 10, pp. 3834-3847, Oct 2021.
\bibitem{b6} D. Liu, Z. Chen, S. Liu and F. Wu, “Deep learning-based technology in responses to the joint call for proposals on video compression with capability beyond HEVC,” \emph{IEEE Transactions on Circuits and Systems for Video Technology}, vol. 30, no. 5, pp. 1267-1280, May 2020.
\bibitem{b7} J. Li, B. Li, J. Xu, R. Xiong, and W. Gao, “Fully connected networkbased intra prediction for image coding,” \emph{IEEE Transactions on Image Processing}, vol. 27, no. 7, pp. 3236–3247, Jul 2018.
\bibitem{b9}W. Cui, T. Zhang, S. Zhang, F. Jiang, W. Zuo, Z. Wan, and D. Zhao, “Convolutional neural networks based intra prediction for HEVC,” in \emph{Proc. Data Compression Conf. (DCC)}, 2017, pp. 436–436.
\bibitem{b10} Y. Wang, X. Fan, S. Liu, D. Zhao and W. Gao, “Multi-scale Convolutional Neural Network Based Intra Prediction for Video Coding,” \emph{IEEE Transactions on Circuits and Systems for Video Technology}, vol. 30, no. 7, pp. 1803-1815, Jul 2020.
\bibitem{b11} J. Pfaff, P. Helle, D. Maniry, S. Kaltenstadler, W. Samek, H. Schwarz, D. Marpe, T. Wiegand, “Neural network based intra prediction for video coding,” in \emph{Proc. Applications of Digital Image Processing XLI}, 2018.
\bibitem{b12} J. Pfaff, B. Stallenberger, M. Schäfer, P. Merkle, P. Helle, T. Hinz, H. Schwarz, D. Marpe, T. Wiegand (HHI), “CE3: Affine linear weighted intra prediction (CE3–4.1 CE3–4.2),”  document JVET-N0217, Geneva, Mar 2019. 
\bibitem{b13} P. Helle, J. Pfaff, M. Schäfer, R. Rischke, H. Schwarz, D. Marpe, T. Wiegand, “Intra picture prediction for video coding with neural networks,” in \emph{Proc. Data Compression Conf. (DCC)}, 2019, pp. 448-457.
\bibitem{TMM} H. Sun, Z. Cheng, M. Takeuchi and J. Katto, “Enhanced Intra Prediction for Video Coding by Using Multiple Neural Networks,” \emph{IEEE Transactions on Multimedia}, vol. 22, no. 11, pp. 2764-2779, Nov 2020.
\bibitem{PSRNN} Hu Y \emph{et al.}, “Progressive spatial recurrent neural network for intra prediction,” \emph{IEEE Transactions on Multimedia}, vol. 21, no. 12, pp. 3024-3037, Dec 2019.
\bibitem{b14} N. Yan, D. Liu, H. Li, B. Li, L. Li, and F. Wu, “Convolutional neural network-based fractional-pixel motion compensation,” \emph{IEEE Transactions on Circuits and Systems for Video Technology}, vol. 29, no. 3, pp. 840-853, Mar 2019.
\bibitem{b15} Z. Zhao, S. Wang, S. Wang, X. Zhang, S. Ma, and J. Yang, “Enhanced bi-prediction with convolutional neural network for high efficiency video coding,” \emph{IEEE Transactions on Circuits and Systems for Video Technology}, vol. 29, no. 11, pp. 3291-3301, Nov 2019.
\bibitem{b16}C. Jia, S. Wang, X. Zhang, S. Wang, J. Liu, S. Pu and S. Ma, “Content-Aware Convolutional Neural Network for In-Loop Filtering in High Efficiency Video Coding,” \emph{IEEE Transactions on Image Processing }, vol. 28, no. 7, pp. 3343-3356, Jul 2019.
\bibitem{b17}F. Wu \emph{et al.}, “Description of SDR video coding technology proposal by University of Science and Technology of China, Peking University, Harbin Institute of Technology, and Wuhan University,” document JVET-J0032, San Diego, Apr 2018.
\bibitem{b18} L. Zhou, X. Song, J. Yao, L. Wang, F. Chen, “Convolutional neural network filter (CNNF) for intra frame,” document JVET-I0022, Gwangju, Jan 2018.
\bibitem{b19} C. Ma, D. Liu, X. Peng, L. Li, and F. Wu, “Convolutional neural network-based arithmetic coding for HEVC intra-predicted residues,” \emph{IEEE Transactions on Circuits and Systems for Video Technology}, vol. 30, no. 7, pp. 1901-1916, Jul 2020.
\bibitem{b20} S. Puri, S. Lasserre, and P. Le Callet, “CNN-based transform index prediction in multiple transforms framework to assist entropy coding,” in \emph{Proc. European Signal Processing Conference (EUSIPCO)}, 2017, pp. 798–802.
\bibitem{b21} R. Yang, M. Xu, T. Liu, Z. Wang, and Z. Guan, “Enhancing quality for HEVC compressed videos,” \emph{IEEE Transactions on Circuits and Systems for Video Technology}, vol. 29, no. 7, pp. 2039-2054, Jul 2019.
\bibitem{b22} Y. Dai, D. Liu, and F. Wu, “A convolutional neural network approach for post-processing in HEVC intra coding,” in \emph{Proc. International Conference on Multimedia Modeling}, 2017, pp. 28–39.
\bibitem{b23}Forgy, Edward W. “Cluster analysis of multivariate data: efficiency versus interpretability of classifications.” \emph{Biometrics} vol. 21, pp. 768-780, 1965.
\bibitem{b24}\emph{VVC Test Model (VTM-4.0).} Accessed: Feb. 12, 2019. [Online].\\ Available: https://vcgit.hhi.fraunhofer.de/jvet/VVCSoftware\_VTM/-/tree/VTM-4.0
\bibitem{b25}\emph{HEVC Test Model (HM-16.9).} Accessed: Mar. 22, 2018. [Online].\\ Available: https://hevc.hhi.fraunhofer.de/svn/svn\_HEVCSoftware/tags/HM-16.9/
\bibitem{b26}B. Bross \emph{et al.}, “General Video Coding Technology in Responses to the Joint Call for Proposals on Video Compression With Capability Beyond HEVC,” \emph{IEEE Transactions on Circuits and Systems for Video Technology}, vol. 30, no. 5, pp. 1226-1240, May 2020.
\bibitem{b27} L. Zhao \emph{et al.}, “Wide angular intra prediction for versatile video coding,” in \emph{Proc. Data Compress. Conf. (DCC)}, 2019, pp. 53–62.
\bibitem{b28} A. Said, X. Zhao, M. Karczewicz, J. Chen, and F. Zou, “Position dependent prediction combination for intra-frame video coding,” in \emph{Proc. IEEE Int. Conf. Image Process. (ICIP)}, 2016, pp. 534–538.
\bibitem{b29}B. Bross \emph{et al.}, “Multiple reference line intra prediction,” document JVET-L0283, Macao, Oct 2018.
\bibitem{b30}M. Karczewicz, E. Alshina. “JVET AHG report: Tool evaluation (AHG1),” document JVET-C0001, Geneva, May 2016.
\bibitem{b31}J. Li, B. Li, J. Xu and R. Xiong, “Efficient Multiple-Line-Based Intra Prediction for HEVC,” \emph{IEEE Transactions on Circuits and Systems for Video Technology}, vol. 28, no. 4, pp. 947-957, Apr 2018.
\bibitem{b32}B. Bross, W.-J. Han, J.-R. Ohm, G. J. Sullivan, Y.-K. Wang, and T. Wiegand, “High Efficiency Video Coding (HEVC) Text Specification Draft 10,” document JCTVC-L1003, Geneva, Jan 2013.
\bibitem{b33}K, He, X. Zhang, S. Ren, J. Sun, “Delving deep into rectifiers: Surpassing human-level performance on imagenet classification.” in \emph{Proc. IEEE International Conference on Computer Vision (ICCV)}, 2015, pp. 1026-1034.
\bibitem{b34}Y. Linde, A. Buzo, R. Gray. “An algorithm for vector quantizer design,” \emph{IEEE Transactions on communications}, vol. 28, no. 1, pp. 84-95, Jan 1980,
\bibitem{b35}Q. Ning. “On the momentum term in gradient descent learning algorithms,” \emph{Neural networks}, Volume 12, Issue 1, pp. 145-151, Jan 1999.
\bibitem{b37}G. J. Sullivan and T. Wiegand, “Rate-distortion optimization for video compression,” \emph{IEEE Signal Processing Magazine}, vol. 15, no. 6, pp. 74–90, Nov 1998.
\bibitem{ISP}S. De-Luxán-Hernández, V. George, J. Ma, T. Nguyen, H. Schwarz, D. Marpe, T. Wiegand (HHI), “CE3: Intra Sub-Partitions Coding Mode (Tests 1.1.1 and 1.1.2)”, document JVET-M0102, Marrakech, Jan 2019.
\bibitem{b38}Marpe. D, Schwarz. H, Wiegand. T, “Context-based adaptive binary arithmetic coding in the H. 264/AVC video compression standard”, \emph{IEEE Transactions on circuits and systems for video technology}, vol. 13, no. 7, pp. 620-636, Jul 2003.
\bibitem{b39}Boyce J, Suehring K, Li X. “JVET common test conditions and software reference configurations[J]”, document JVET-J1010, San Diego, Apr 2018.
\bibitem{b40}K. Sharman and K. Suehring,“Common Test Conditions”, document JCTVC-Z1100, Geneva, Jan 2017.
\bibitem{b41}G. Bj$\phi$ntegaard. “Improvements of the BD-PSNR mode,” document ITU-T SG16 Q.6, VCEG-AI11, Berlin, Jul 2008.
\bibitem{b42}K. Wilson and N. Snavely, “Robust global translations with 1DSfM,” in \emph{Proc. European Conference on Computer Vision}, 2014, pp. 61–75.
\bibitem{b43}A. Paszke \emph{et al.}, “PyTorch: An Imperative Style, High-Performance Deep Learning Library,” in \emph{Proc. Advances in neural information processing systems}, 2019, pp 8026-8037.
\end{thebibliography}
\end{document}